\documentclass[a4paper,10pt]{cpc-hepnp}
\usepackage{CJK,upgreek,fancyhdr}
\usepackage{subfigure}
\usepackage{multicol}
\usepackage{booktabs}
\usepackage{amssymb,bm,mathrsfs,bbm,amscd}
\usepackage[tbtags]{amsmath}
\usepackage{lastpage}
\usepackage{graphicx}
\usepackage{multirow}

\DeclareGraphicsExtensions{.pdf,.jpeg,.png}

\usepackage{epstopdf}
\usepackage[english]{babel}
\usepackage[colorlinks=true,urlcolor=blue,linkcolor=blue,anchorcolor=blue,citecolor=blue,pdfpagemode=FullScreen,,setpagesize=off,pdfborder={0 0 0}]{hyperref}
\usepackage{overpic}
\usepackage{lineno}
\usepackage{color}

\newcommand {\br}[1]{\mathcal{B}(#1)}

\newcommand {\ie}       {\emph{i}.\emph{e}.}

\def \Lcm{\overline{\Lambda}{}_{c}^{-}}
\def \Lcp{\Lambda_{c}^{+}}
\def \Xcp{\Xi_{c}^{+}}
\def \Xcz{\Xi_{c}^{0}}

\def \Ocz{\Omega_{c}^{0}}
\def \Lz{\Lambda}
\def \Kstarbzz{\overline{K}{}_{0}^{*}}
\def \Kstarb{\overline{K}{}^{*}}
\def \Kstarbtwo{\overline{K}{}_{2}^{*}}

\def \pip{\pi^+}

\def \pim{\pi^-}

\def \ee{e^+e^-}
\def \mumu{\mu^+\mu^-}

\def \kp{K^+}
\def \km{K^-}

\def \Ks {K_{S}^{0}}

\def \BR{\mathcal{B}}

\def \Bc{\mathcal{B}_c}
\def \ipb  {\,\mbox{pb$^{-1}$}}
\def \ifb  {\,\mbox{fb$^{-1}$}}

\def \mbc {M_{\rm{BC}}}

\def \miss2{M_{\rm miss}^{2}}

\begin{document}
\begin{CJK*}{UTF8}{gkai}

  \fancyhead[c]{\small Chinese Physics C~~~Vol. 50, No. 2 (2026) 022002}
  \fancyfoot[C]{\small 022002-\thepage}

  \footnotetext[0]{Received xxxx June xxxx}

  \title{Experimental overview on the charmed baryon decays
    \thanks{
      The authors thank Fusheng Yu for useful discussions. The authors are supported in part by National Natural Science Foundation of China under contracts Nos. 12221005 and 12422504; CAS Project for Young Scientists in Basic Research under contract No. YSBR-117; Fundamental Research Funds for the Central Universities, Lanzhou University under contracts Nos. lzujbky-2025-ytB01, lzujbky-2023-stlt01 and lzujbky-2023-it32, University of Chinese Academy of Sciences.
    }}
  \maketitle

  \vspace{-20pt}

  \begin{center}
    {\large Pei-Rong Li (李培荣)}$^{a}$~\footnote{E-mail: prli@lzu.edu.cn},
    {\large Xiao-Rui Lyu (吕晓睿)}$^b$~\footnote{E-mail: xiaorui@ucas.ac.cn} \\
    {\large Yangheng Zheng (郑阳恒)}$^b$~\footnote{E-mail: zhengyh@ucas.ac.cn} \\
    \vspace{0.3cm}
    $^{a}$ Lanzhou University, Lanzhou 730000, People's Republic of China\\
    $^{b}$ University of Chinese Academy of Sciences, Beijing 100049, People's Republic of China
  \end{center}


  \begin{abstract}
  The charmed baryon was first observed experimentally in 1975, one year after the charm quark's confirmation via the discovery of the $J/\psi$ particle. Studying charmed baryon decays provides a pathway to investigate both strong and weak interactions, leveraging the weak decays of the embedded charm quark. However, for approximately three decades following its discovery, experimental knowledge of charmed baryons remained significantly limited compared to those of the hidden-charm $\psi$ mesons and open-charm $D_{(s)}$ mesons. This situation changed markedly starting in 2014, when dedicated data collection for charmed baryons commenced at BESIII. In this article, we review the experimental progress achieved since 2014 in understanding the weak decays of the charmed baryons.
  \end{abstract}




  \section{Introduction}
\begin{figure}[htbp]
  \centering
  \includegraphics[width=0.5\textwidth]{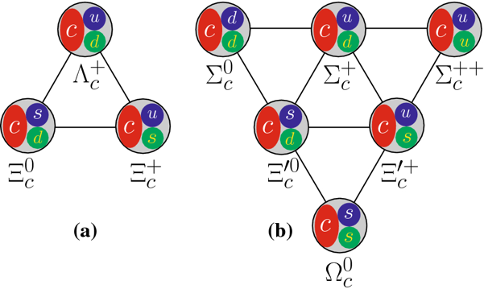}
  \caption{Family of antitriplet and sextet ground-state charmed baryons.
    \label{fig:cbaryon_family}}
\end{figure}

  Seeking of the fundamental composition of matter dates back over two thousand years, and today, the widely accepted theory is the Standard Model (SM) of particle physics. The modern SM of particle physics describes the composition of matter around us using 17 fundamental particles and three interactions: the strong force, the weak force, and electromagnetic interaction. Among these fundamental particles, quarks form hadrons through the strong force. Protons and neutrons are the most familiar hadrons, belonging to a class of hadrons called baryons, composed of three quarks. A baryon containing at least one charm quark is called a charmed baryon.
In the spectrum of charmed baryons (also denoted as $\Bc$ in this paper), the ground-state charmed baryons containing one charm quark form an antitriplet and a sextet as shown in Fig.~\ref{fig:cbaryon_family}. These baryons are not as stable as protons and quickly decay into lighter, longer-lived particles. The antitriplet charmed baryons $\Lcp$($cud$), $\Xcp$($cus$) and $\Xcz$($cds$) can only decay through quark weak decay; in the sextet, only the heaviest $\Ocz$($css$) decays through weak interaction, while $\Sigma_c^{++}$($cuu$), $\Sigma_c^+$($cud$), $\Sigma_c^0$($cdd$) almost exclusively decay through $\pi$ meson strong interaction transitions to $\Lcp$ with lighter mass. The $\Xi_c^{\prime +}$($cus$) and $\Xi_c^{\prime 0}$($cds$) decay to the same charged $\Xcp$ and $\Xcz$ through photon transition, respectively. Thus, the weak decay modes of the four charmed baryons $\Lcp$, $\Xcp$, $\Xcz$ and $\Ocz$ make their lifetimes relatively long, with richer decay modes and more complex interaction mechanisms involved. Experimentally studying these four charmed baryon properties in detail can be used not only to test weak interaction theories but also to test strong interaction mechanisms. It is an important means to precisely test SM and search for new physics.
In addition, most excited states of charmed baryons and bottom baryons mostly finally decay to the ground state charmed baryons. Therefore, accurately measuring the properties of the ground state charmed baryons is of significant physical importance for understanding hadron spectroscopy and testing SM in the bottom sector.

In experiment, the first charmed baryon was discovered in 1975 in the 7-inch low-temperature bubble chamber of the Brookhaven National Laboratory (BNL) through the detection of neutrino beams. Due to the reaction of neutrinos with protons in the detector material, the reaction $v p \rightarrow \mu^- \Lambda \pi^+ \pi^+ \pi^+ \pi^-$ was obtained~\cite{Cazzoli:1975et}, where $\Lambda \pi^+ \pi^+ \pi^- $ comes from $\Sigma_c^{++} \rightarrow \Lambda_c^+ \pi^+ $, $\Lcp \to \Lambda \pip \pip \pim$. Subsequently, in 1976, the  $\Lambda_c^+$ was confirmed in the decays to $\Lambda \pip \pip \pim$ at Fermi National Laboratory (FermiLab) through photon production~\cite{Knapp:1976qw}. In 1980, the MarkII experiment at the Stanford Linear Accelerator Center utilized the process of positron-electron annihilation to firstly measure the production cross-section and mass of the $\Lcp$~\cite{Abrams:1979iu}, which initiated experimental research on the properties of charmed baryons. Similar to the discovery of the $\Sigma_c^{++}$ in BNL~\cite{Cazzoli:1975et}, through neutrino reactions, the $\Sigma_c^{+}$ and $\Sigma_c^{0}$ were later discovered at CERN~\cite{BEBCTSTNeutrino:1980ktj} and at FermiLab~\cite{Voyvodic:1986ug}, respectively.
In 1983 the $\Xcp$ was discovered in an experiment at the CERN SPS hyperon beam~\cite{Biagi:1983en} and in 1989 the $\Xcz$ was observed in $\ee$ annihilations at CLEO~\cite{CLEO:1988yda}. The $\Ocz$ was firstly reported in the WA62 experiment utilizing the SPS charged hyperon beam at CERN~\cite{Biagi:1984mu}. The observation of the $\Xi_c^{\prime +}$ and $\Xi_c^{\prime 0}$ were simultaneously reported at CLEO~\cite{CLEO:1998wvk}.

Since the discoveries of the charmed baryons, there had been many theoretical studies on their properties in the early 1990s. However, the hot period subsequently faded away, since the experimental measurements were retarded~\cite{Asner:2008nq}. 
Since 2014, there have been significant developments in the experimental studies of the charmed baryons from  BESIII, LHCb and BELLE. For example, the systematic studies on the $\Lambda_c^+$ productions and decays at BESIII~\cite{Li:2021iwf} have substantially expanded the experimental data available in partical data group (PDG)~\cite{ParticleDataGroup:2024cfk}.
The masses and lifetimes of the singly charmed baryons have been well determined in experiment~\cite{ParticleDataGroup:2024cfk}. In particular, the lifetime hierarchy was determined by the LHCb collaboration~\cite{LHCb:2018nfa,LHCb:2019ldj,LHCb:2021vll,Cheng:2021vca,LHCb:2025oww} as $\tau_{\Xi_c^+} > \tau_{\Omega_c^0} > \tau_{\Lambda_c^+} > \tau_{\Xi_c^0}$, which changed from the previous order $\tau_{\Xi_c^+}>\tau_{\Lambda_c^+}>\tau_{\Xi_c^0}>\tau_{\Omega_c^0}$ before 2018.
In addition, the spin of the $\Lcp$ is firstly determined to be $1/2$~\cite{BESIII:2020kap}, which is consistent with the theoretical prediction of the quark model.
These advancements sparked renewed theoretical interests in the studies of singly charmed baryons, as it challenges previous expectations and necessitates a deeper understanding of the underlying dynamics. 
A series of reviews on the developments of the theoretical and experimental studies on the charmed baryons can be found in Refs.~\cite{Bianco:2003vb,Asner:2008nq,Cheng:2015iom,Gronau:2018vei,Bianco:2020hzf,Cheng:2021qpd,Li:2021iwf,BESIII:2020nme,BaBar:2014omp}.

\begin{figure}[htbp]
\centering
\subfigure[]{
\includegraphics[width=0.45\textwidth]{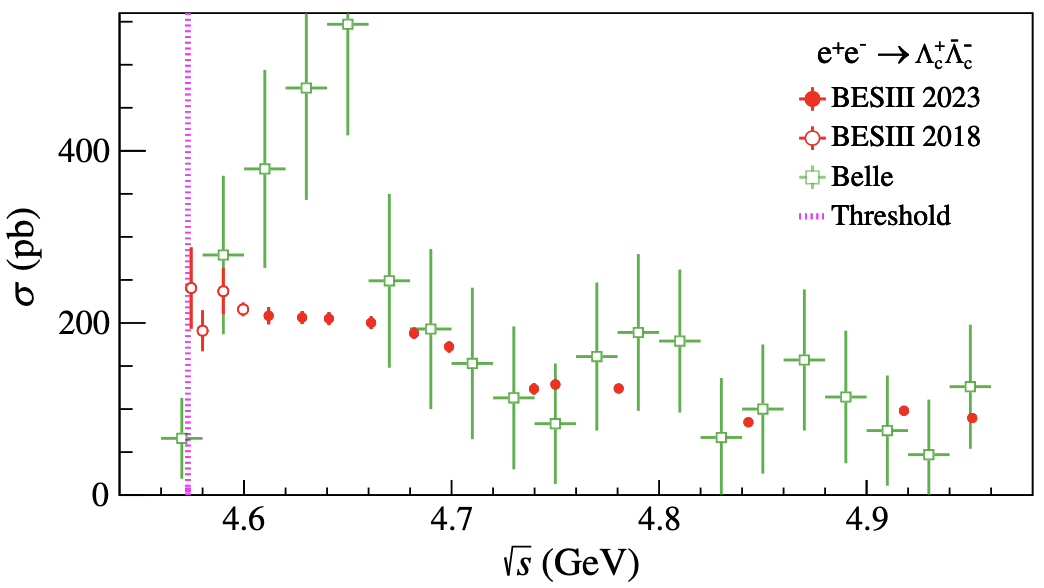}
\label{fig:cross_section}
}
\subfigure[]{
\includegraphics[width=0.45\textwidth]{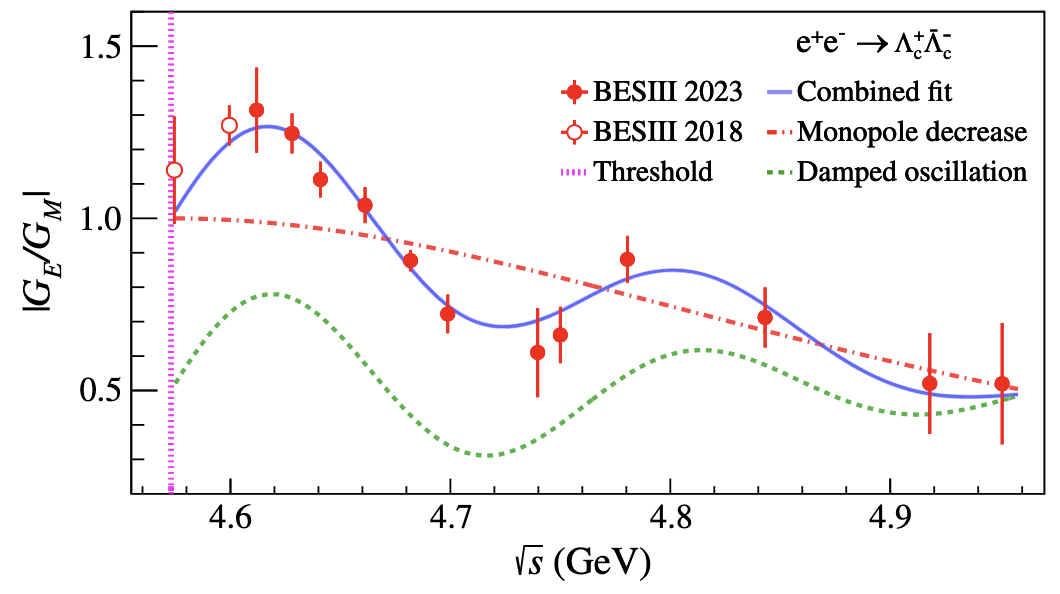}
\label{fig:form_factor}
}
\caption{Plot (a): Distribution of the cross sections of the production $\ee\to\Lcp\Lcm$ measured by BESIII~\cite{BESIII:2023rwv,BESIII:2017kqg} and Belle~\cite{Belle:2008xmh}. Plot (b): Distributions of the extracted effective form factors  measured by BESIII~\cite{BESIII:2023rwv,BESIII:2017kqg}}.
\end{figure}

BESIII accumlated $e^+e^-$ collision data just above the $\Lcp\Lcm$ mass threshold in the energy region between 4.6 GeV and 4.95 GeV with an integrated luminosity of 6.4\ipb~\cite{BESIII:2022ulv}, which consist of about 0.8 millions of $\Lcp\Lcm$ pairs.
As the $\Lcp\Lcm$ pairs are produced through the two-body channel with no additional particles, one can employ a double-tag (DT) technique, pioneered in the Mark III experiment~\cite{MARK-III:1985hbd}.
Namely, full reconstruction of a $\Lcm$ on one side of tagged events effectively provides a ``beam"  of $\Lcp$ particles with known four-momentum on the other side.
The tag yield, which provides the normalization for the BF measurement, is extracted from the distribution of beam-constrained mass $\mbc c^2 = \sqrt{(\sqrt{s}/2)^2 - |\vec{p}_{\rm tag} c|^2}$, where $\vec{p}_{\rm tag}$ is the three-momentum of the tagging $\Lcm$ candidate and $\sqrt{s}$ is the center-of-mass energy of the $\ee$ system.
When a tagged $\Lcp$ decays
to $\Lambda e^+$ and an electron neutrino, $\Lambda e^+ \nu_e$, the mass of the (missing) zero-mass neutrino
can be inferred from energy-momentum conservation. This tagging technique, which obviates the need for knowledge of the luminosity or the production cross section, is a powerful tool for charmed particle
decay measurements that is only possible in the near-threshold experiments.
Based on the tagging technique, BESIII accurately measured the cross sections of the $\Lcp\Lcm$ production~\cite{BESIII:2017kqg, BESIII:2023rwv} from mass threshold to 4.95 GeV , as shown in Fig.~\ref{fig:cross_section}. Here the cross sections at BESIII indicate no enhancement around the $Y(4630)$ resonance, which is different from Belle~\cite{Belle:2008xmh}.  The great precisions of the cross sections at BESIII allow for extraction of the effective form factors for the first time, as illustrated in Fig.~\ref{fig:form_factor}, which reveals an oscillation feature as a function of energy.
Furthermore, the charmed baryon pairs are produced via $e^+e^-$ annihilation through a virtual photon ($J^{PC} = 1^{--}$), {\it e.g.,}  in the process $e^+e^- \to \gamma^* \to \Lcp \Lcm$,  the wave function of the $\Lambda^+_c \bar{\Lambda}^-_c$ is analogous to that of photons in a spin-triplet state with odd charge parity $C=-$, and the $\Lambda^+_c \bar{\Lambda}^-_c$ pair are in a quantum-entangled state, which allows for unique probes of the structure of decay amplitudes,
as well as the polarization and $CP$ violation in $\Lambda^+_c$ decays~\cite{Kang:2010td, Li:2021iwf}.

During its first operational period (RUN1, 2010$-$2012), the LHCb experiment collected data at collision energies of 7 TeV and 8 TeV with a total integrated luminosity of 3.2$\ifb$. The second operational period (RUN2, 2015-2018) accumulated 5.9$\ifb$ of data at 13 TeV.  Within the acceptance of the LHCb detector in these collision energy ranges, charmed baryons are produced with cross-sections on the order of hundreds of microbarns, resulting in billions of charmed baryons in the dataset.
Due to very high multiplicities in the proton-proton collisions, LHCb is extremely avdantagous in detecting the charmed baryons with purely charged final states.

Similar to LHCb, the charmed baryons at BELLE (II) can be produced directly through the $e^+e^-$ annihilation process and secondarily from the $B$ meson decays.
So far, BELLE has collected about 1~ab$^{-1}$ of data at $\Upsilon(4S)$.
As the corresponding cross section of the direct charm production is about 1.3 nb at $\sqrt{s}=10.58$ GeV, \ie, the $\Upsilon(4S)$ resonance, which is at same order of the bottom production cross section, most of studies on the charmed baryons are based on the direct produciton process in order to sustain the high statistics.
However, the backgrounds from continuum processes are inreducible, which makes the analyses of the charmed baryons subject to large systematic uncertainties.

  \section{Semi-leptonic decays}
  The semi-leptonic (SL) decay of the charmed baryon provides unique insights into the fundamental mechanism of strong and electroweak interactions, serving as a testbed for investigating non-perturbative quantum chromodynamics (QCD) effects and constraining the Cabibbo-Kobayashi-Maskawa (CKM) matrix parameters.
In 1990s, only quite a few SL decays of the charmed baryons are reported in experiment.
The first SL mode observed is $\Lcp\to \Lambda e^+ \nu_e$ in the ARGUS experiment~\cite{ARGUS:1991bvx} and later confirmed by CLEO~\cite{CLEO:1994lyw}.
The mode $\Xcz \to \Xi^- \ell^+ \nu_\ell$ ($\ell=e$ or $\mu$) was firstly reported by ARGUS~\cite{ARGUS:1992jnv}.
CLEO confirmed $\Xcz \to \Xi^- e^+ \nu_e$ and observed $\Xcp \to \Xi^0 e^+ \nu_e$~\cite{CLEO:1994aud}.
Later, CLEO observed the SL decay $\Ocz \to \Omega^- e^+ \nu_e$~\cite{CLEO:2002imi}.
So far, the SL decays of all the charmed baryons have been observed in experiment, including $\Lcp$, $\Xcz$, $\Xcp$ and $\Ocz$.
However, in these early experimental studies, the product of the cross section $\sigma(\ee\to \Bc X)$ and the $\Bc$ SL BF at $B\bar{B}$ threshold energies are measured with poor precisions.
Hence, only relative BFs were directly measured and no straightforward access to the absolute SL BFs were available.

\begin{table*}[tp]
    \caption{Determined BFs for SL decays of the charmed baryons. The BFs labelled with $^\dagger$ are products of the directly determined relative BFs and the BFs for the normalization channels, which are quoted from the latest PDG~\cite{ParticleDataGroup:2024cfk} such as $\br{\Lcp\to pK^-\pi^+}$ = $(6.24\pm0.28)\%$, $\br{\Xcz\to \Xi^- \pi^+}$ = $(1.43\pm0.27)\%$ and $\br{\Xcp\to \Xi^- \pip \pip}$ = $(2.9\pm1.3)\%$.
        For the $\Ocz$ SL decays, the ratio of the BF relative to $\BR(\Ocz\to\Omega^-\pi^+)$ is given, as no absolute BFs are determined yet. Relative precisions are given in parentheses.
        \label{tab:BcSL}}
    \begin{center}
        \footnotesize
        \begin{tabular}{lcc|lcc}
            \hline\hline
            $\Lcp$ Mode                                     & BF($\times 10^{-3}$)                   & Experiment                           & $\Lcp$ Mode                                        & BF($\times 10^{-3}$)  & Experiment                      \\ \hline
            \multirow{4}{*}{$\Lcp\to\Lambda e^+ \nu_e$}     & 23.7$\pm$5.1(37\%)$^\dagger$           & ARGUS(1991)\cite{ARGUS:1991bvx}
                                                            & $\Lcp\to p K^- e^+ \nu_e$              & $0.88\pm0.18$(20\%)                  & BESIII(2022)\cite{BESIII:2022qaf}                                                                            \\ \cline{4-6}
                                                            & 26.8$\pm$5.1(19\%)$^\dagger$           & CLEO(1994)\cite{CLEO:1994lyw}
                                                            & $\Lcp\to \Lambda(1405) e^+ \nu_e$,     & \multirow{2}{*}{0.42$\pm$0.19(45\%)} & \multirow{2}{*}{BESIII(2022)\cite{BESIII:2022qaf}}                                                           \\
                                                            & 36.3$\pm$4.3(12\%)                     & BESIII(2015)\cite{BESIII:2015ysy}
                                                            & $\Lambda(1405)\to p K^-$               &                                      &                                                                                                              \\  \cline{4-6}
                                                            & 35.6$\pm$1.3(3.6\%)                    & BESIII(2022)\cite{BESIII:2022ysa}
                                                            & $\Lcp\to\Lambda(1520) e^+ \nu_e$       & $1.0\pm0.5$(50\%)                    & BESIII(2022)\cite{BESIII:2022qaf}                                                                            \\ \cline{1-3}
            \multirow{2}{*}{$\Lcp\to\Lambda \mu^+ \nu_\mu$} & 34.9$\pm$5.3(15\%)                     & BESIII(2017)\cite{BESIII:2016ffj}
                                                            & $\Lcp\to p\Ks\pim e^+ \nu_e$           & $<0.33$                              & BESIII(2023)\cite{BESIII:2023jem}                                                                            \\
                                                            & 34.8$\pm$1.7(4.9\%)                    & BESIII(2023)\cite{BESIII:2023jxv}
                                                            & $\Lcp\to\Lambda \pip \pim e^+ \nu_e$   & $<0.39$                              & BESIII(2023)\cite{BESIII:2023jem}                                                                            \\ \cline{1-3}
            \multirow{2}{*}{$\Lcp\to e^+ X$}                & $39.5\pm3.5$(8.9\%)                    & BESIII(2018)\cite{BESIII:2018mug}
                                                            & $\Lcp\to n e^+ \nu_e$                  & $3.57\pm0.37$ (10\%)                 & BESIII(2025)\cite{BESIII:2024mgg}                                                                            \\
                                                            & $40.6\pm1.3$(3.2\%)                    & BESIII(2023)\cite{BESIII:2022cmg}    &                                                    &                       &                                 \\ \hline \hline
            $\Xi_c$ Mode                                    & BF($\times 10^{-3}$)                   & Experiment                           & $\Xi_c$ Mode                                       & BF($\times 10^{-3}$)  & Experiment                      \\ \hline
            \multirow{4}{*}{$\Xcz \to \Xi^- e^+ \nu_e$}     & $13.7\pm7.7$(56\%)$^\dagger$           & ARGUS(1993)\cite{ARGUS:1992jnv}
                                                            & $\Xi_c^0 \to \Xi^- \mu^+ \nu_\mu$      & $10.1\pm2.1$(21\%)$^\dagger$         & Belle(2021)\cite{Belle:2021crz}                                                                              \\
                                                            & $44.3^{+16.6}_{-17.8}$(40\%)$^\dagger$ & CLEO(1995)\cite{CLEO:1994aud}
                                                            & $\Xi_c^+ \to \Xi^0 e^+ \nu_e$          & $67\pm39$(58\%)$^\dagger$            & CLEO(1995)\cite{CLEO:1994aud}                                                                                \\
                                                            & $19.7\pm5.3$(27\%)$^\dagger$           & ALICE(2021)\cite{ALICE:2021bli}      &                                                    &                       &                                 \\
                                                            & $10.4\pm2.1$(20\%)$^\dagger$           & Belle(2021)\cite{Belle:2021crz}      &                                                    &                       &                                 \\ \hline \hline
            $\Ocz$ Mode                                     & Ratio                                  & Experiment                           & $\Ocz$ Mode                                        & Ratio                 & Experiment                      \\ \hline
            \multirow{2}{*}{$\Ocz \to \Omega^- e^+ \nu_e$}  & $2.4\pm1.1$(47\%)                      & CLEO(2002)\cite{CLEO:2002imi}        & $\Ocz \to \Omega^- \mu^+ \nu_\mu$                  & $1.94 \pm 0.21$(11\%) & Belle(2022)\cite{Belle:2021dgc} \\
                                                            & $1.98\pm0.15$(7.7\%)                   & Belle(2022)\cite{Belle:2021dgc}      &                                                    &                       &                                 \\ \hline  \hline
        \end{tabular}
    \end{center}
\end{table*}

\begin{figure}[tp]
    \centering
    \includegraphics[width=0.35\textwidth]{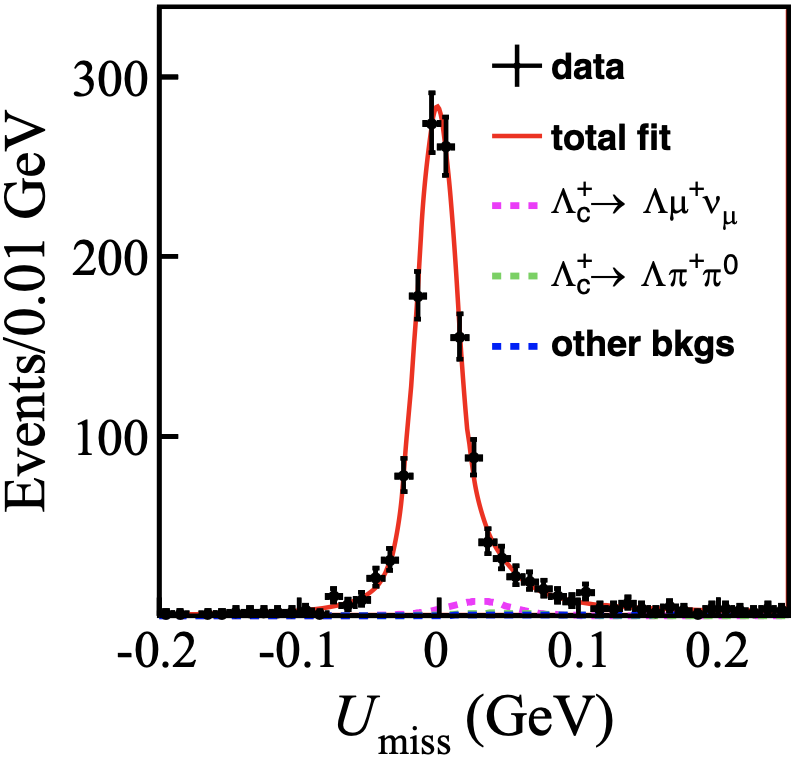} \hspace{0.1cm}
    \includegraphics[width=0.52\textwidth]{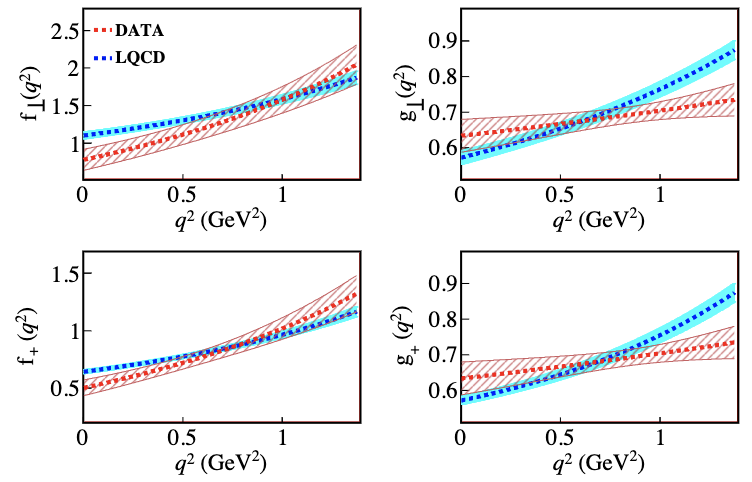}
    \caption{(left) The missing mass distribution for the neutrino signals in $\Lcp\to \Lambda e^+ \nu_e$. (right) Comparisons of the determined form factors in $\Lcp\to \Lambda \ell^+ \nu_\ell$  with LQCD calculation.
        \label{fig:SL}}
\end{figure}

Based on the unique threshold data collected at center-of-mass energies right above the total mass of the charmed baryon pair, BESIII is capable of
determining the absolute BFs using the double-tag and missing-mass technique~\cite{Li:2021iwf}, as shown in the left plot of Fig.~\ref{fig:SL}.
The first absolute measurement of the SL BF was realized in fitting to the missing mass distribution of the $\Lcp\to \Lambda e^+ \nu_e$ decay~\cite{BESIII:2015ysy} by analyzing the $\ee$ collision dataset of 587\ipb at 4.6 GeV at BESIII~\cite{BESIII:2015ysy}, where $\BR(\Lcp\to \Lambda e^+ \nu_e)$ was given as $(3.63\pm0.38\pm0.20)$\%.
The precision is significantly improved over previous indirect measurements and imposes stringent  constraints on various phenomenological models. Notably, this spurs the first Lattice QCD (LQCD) calculation for the SL decays of the charmed baryons SL decay~\cite{Meinel:2016dqj}, the outcomes of which are consistent with the BESIII result.
With the same dataset, the decay of $\Lcp\to \Lambda \mu^+ \nu_\mu$ is observed, which is the first semi-muonic decay of the charmed baryon seen in experiment, and the absolute BF were determined as $(3.49\pm0.46\pm0.27)\%$~\cite{BESIII:2016ffj}, which is consistent with the semi-electronic mode within the lepton flavor universality.
Since then, BESIII carried out a series of studies on the SL decays of the $\Lcp$ after a larger $\Lcp$ threshold data between 4.61 GeV and 4.95 GeV with an integrated luminosity of 5.9\ifb was collected~\cite{BESIII:2022ulv}.
The second SL decay of the $\Lcp\to p K^- e^+ \nu_e$, is observed for the first time and the corresponding absolute BF is measured~\cite{BESIII:2022qaf}. In this mode, the first evidence for the intermediate process $\Lcp \to \Lambda(1520) e^+ \nu_e$ is uncovered. Furthermore, searches were conducted for the decays such as $\Lcp \to \Lambda \pip\pim e^+ \nu_e$ and $p K_S^0 \pip e^+ \nu_e$~\cite{BESIII:2023jem}.

The above results concern Cabibbo-favored (CF) SL decays.
The only observed Cabibbo-suppressed (CS) SL decay is $\Lambda_c^+ \to n \ell^+ \nu_{\ell}$, reported at BESIII with a significance of $5.2\sigma$~\cite{BESIII:2024mgg}. 
In this channel, as there are two missing neutral particles (neutron and neutrino) in the final states, the advanced machine learning approach is adopted in discriminating the shower patterns in the electromagnetic calorimeter induced by the neutron from the $\Lambda\to p \pi^0$ backgrounds. The BF is measured as $(3.57\pm0.37\pm0.14)\%$, which provides a new calibration for various QCD-inspired phenomenological models and LQCD calculation~\cite{Meinel:2017ggx}. Additionally, by quoting the $\Lcp$ lifetime and the form factors obtained in the LQCD calculation~\cite{Meinel:2017ggx}, the CKM matrix element $|V_{cd}|$ is extracted via charmed baryon decays for the first time. The obtained $|V_{cd}|$ is $0.208\pm0.011\pm0.007$, which is consistent with the word average result~\cite{ParticleDataGroup:2024cfk}. This work highlights the first CS SL decay of the charmed baryon seen in experiment and showcases the power of the application of the machine learning techniques in experimental particle physics.

BESIII also determined the inclusive SL decay rate of $\Lcp \to e^+ X$~\cite{BESIII:2018mug,BESIII:2022cmg}, by adopting the double-tag method, which is given as $4.06 \pm 0.10\pm 0.09$\%.
Then, by quoting the $\Lcp$ lifetime, the decay width of the inclusive SL decay is derived as $\Gamma(\Lambda_c^+ \to Xe^{+}\nu_e) = (2.006 \pm 0.073) \times 10^{11} \, \text{s}^{-1}$, which provides strong constraints on the theoretical models on the lifetimes of the charmed bayrons.
Table~\ref{tab:BcSL} lists the determined BFs for the SL decays.
Assumming $\br{\Lcp\to p K^- e^+ \nu}=\br{\Lcp\to n \bar{K}^0 e^+ \nu}$, the total BF of known exclusive SL decay channels is calculated to be $(41\pm 1)\times 10^{-3}$ as given in Table~\ref{tab:BcSL}.
When compared to the inclusive SL decay rate $(40.6\pm1.3)\times 10^{-3}$, it implies that the unobserved (``missing") SL decay modes contribute at the level of
$\mathcal{O}(10^{-3})$ to the total decay width of the $\Lcp$.

The SL decay rates of depend critically on the form factors that describe the transition-matrix elements between the initial and final baryon states.
Along with the progress of the measurement of the BFs for the exclusive semi-leptonic decays, there have been great progress in predicting the form factors in theory.
Besides the LQCD calculations~\cite{Meinel:2016dqj,Meinel:2017ggx,Zhang:2021oja,Meinel:2021grq,Meinel:2021mdj,Farrell:2025gis}, 
various theoretical models have been adopted to evaluate these form factors, including the nonrelativistic quark model~\cite{Perez-Marcial:1989sch,Singleton:1990ye,Cheng:1995fe,Pervin:2005ve,Hussain:2017lir}, 
MIT bag model~\cite{Perez-Marcial:1989sch,Geng:2020fng}, relativistic quark models~\cite{Ivanov:1996fj,Gutsche:2014zna,Faustov:2016yza}, light-front quark model~\cite{Luo:1998wg,Zhao:2018zcb,Geng:2020gjh}, QCD sum rules~\cite{MarquesdeCarvalho:1999bqs,Huang:2006ny,Azizi:2011mw,Li:2016qai,Zhang:2023nxl}.
In addition, the SL decays provide clean testes of the SU(3) symmetry in charmed baryon decays~\cite{Geng:2017mxn,Lu:2016ogy,Geng:2019bfz,He:2021qnc}.
Before BESIII, only relative form factors in $\Lcp\to \Lambda e^+ \nu_e$ were firstly studied at CLEO~\cite{CLEO:1995cnb}. However, it is crucial to obtain absolute form factors and improve the precisions to test different theoretical models.
Taking advantage of the large dataset collected at the $\Lcp$ threshold, BESIII has carried out the first measurement of the absolute form factors for the SL decays of charmed baryons.
Four-dimensional fits based on helicity amplitudes were conducted to extract the absolute form factors in the decays $\Lcp\to \Lambda e^+ \nu_e$ and $\Lcp\to \Lambda \mu^+ \nu_\mu$~\cite{BESIII:2022ysa,BESIII:2023jxv}. As can be seen in the right plots in Fig.~\ref{fig:SL}, the comparsions to the LQCD calucations indicate some deviations around 2$\sigma$ level and the lepton flavor universality is verified in different $q^2$ regions.
In addition, the precisions of $\BR(\Lcp\to \Lambda \ell^+ \nu_\ell)$  are improved to be less than 5\%, as listed in Table~\ref{tab:BcSL}.

For the $\Xi_c$ SL decays, recently BELLE and ALICE updated the decay rates of $\Xcz\to \Xi^- \ell^+ \nu_\ell$~\cite{Belle:2021crz,ALICE:2021bli} by taking the decay $\Xcz \to \Xi^- \pi^+$ as reference channel and the rate between the semi-muonic and semi-electronic BFs is obtained as $1.03\pm0.05\pm0.07$.
According the SU(3) asymmetry, with inputs of the recent BESIII measurement of $\br{\Lcp\to\Lambda e^{+}\nu_{e}}$ and $\br{\Lcp\to ne^{+}\nu_{e}}$, at leading order, $\br{\Xcz\to \Xi^- e^+ \nu_e}$ is expected to be around 4 to 5\%~\cite{Farrell:2025gis}, which is significantly larger than the current experimental results. Furthermore, the LQCD calculations predict quite different different $\br{\Xcz\to \Xi^- e^+ \nu_e}$ as $(2.38\pm0.45)\%$~\cite{Zhang:2021oja} and $(3.58\pm0.12)\%$~\cite{Farrell:2025gis}. This requires further experimental studies to clarify the discrepancies.

As for the $\Ocz$, BELLE recently presented the observation of the muonic decay $\Ocz\to \Omega^- \mu^+ \nu_\mu$ and improved the precisions on the ratios of the BFs for $\Ocz\to \Omega^- \ell^+ \nu_\ell$ compared to the reference mode $\Ocz\to \Omega^- \pi^+$. In addition, The ratio of $\BR(\Ocz\to \Omega^- \mu^+ \nu_\mu)/\BR(\Ocz\to \Omega^- e^+ \nu_e)$ is determined to be $1.02\pm0.10\pm0.02$.
All the testes of lepton flavor universality based on $\Xi_c$ and $\Omega_c^0$ SL decays comply with the SM expectation, as the current precisions are still limited, as given in Table~\ref{tab:BcSL}.
Especaially, there has been no absolute BFs available for the $\Ocz$ decays.
Furthermore, oppposite to the $\Lcp$ SL decays, absolute form factors in the $\Xi_c$ and $\Ocz$ SL ecays are missing in the current experimental studies, which would supply crucial test on different theorectical models and LQCD calculations.

  \section{Hadronic decays}
 \setlength{\footskip}{11.3pt} 
The study of charmed baryon hadronic decays provides crucial insights into the interplay between weak and strong interactions, allowing for precise tests of the Standard Model and exploration of potential new physics.
Therefore, comprehensive experimental measurements of various $\Lcp$ hadronic decays play an important role in improving different theoretical calculations~\cite{Cheng:2015iom} and developing the QCD methodology in handling non-perturbative effects.
Before 2014, only about 40\% of the total $\Lcp$ hadronic decay rate had been measured and many modes were not identified, such as those with final state neutrons and $K_L^0$. For $\Xi_c$ and $\Ocz$, studies on the hadronic decays were even more limited, with only a few decay modes observed and measured.
In recent years, a significant number of measurements and discoveries in the $\Lcp$, $\Xi_c$ and $\Ocz$ decays, have been made particularly in the BESIII, Belle (II) and LHCb experiments, refining our understanding of strong dynamics within charmed baryons.
Given the large uncertainties of theoretical treatement of non-perturbative QCD effects in the charmed baryon sector, comprehensive experimental measurements are essential to constrain phenomenological models and guide theoretical advances~\cite{Korner:1992wi,Uppal:1994pt,Zenczykowski:1993hw,Chau:1995gk,Sharma:1996sc,Kohara:1998jm,Ivanov:1997ra,Asner:2008nq}.

\subsection{\texorpdfstring{$\Lcp$ decays}{Lcp decays}}
\subsubsection{BF measurement}

The lightest charmed baryon $\Lcp$, with quark configuration $udc$, serves as the cornerstone of  charmed baryon spectroscopy.
Measurements of various hadronic decays, including two-body and multi-body modes, have been performed, involving different baryonic final states such as $p/n/\Delta$, $\Lambda$, $\Sigma$ and $\Xi$.
These $\Lcp$ BF results from 2014 to 2025 are systematically compiled in Tables~\ref{tab:LcHad_CFBF} and~\ref{tab:LcHad_CSBF}, corresponding to CF and CS decays, respectively.

\vspace{0.2cm}
$\bullet$ CF decays
\vspace{0.2cm}

The decay $\Lcp \rightarrow p K^-\pi^+$ is the golden channel in many studies on the $\Lcp$ properties, as it has the largest BF among all $\Lcp$ hadronic decays.
For instance, in the hadron collider and $B$ factory, most of $\Lcp$ decay BFs were obtained by measuring their ratios to the reference mode $\Lcp \rightarrow p K^-\pi^+$.
It also serves as the high-efficient reconstruction mode of the $\Lcp$ baryon in the hadron collider experiments.
However, the previously determined average BF, $\BR({\Lcp \rightarrow p K^-\pi^+})=(5.0\pm1.3)\%$, had a large uncertainty due to model assumption on inclusive $\Lcp$ production around the $B\bar{B}$ energy in these measurements~\cite{CLEO:2000rnc}.

To resolve this issue, BESIII has collected a data sample of 587 pb$^{-1}$ at 4.6 GeV in 2014, which is just above the $\Lcp\Lcm$ production threshold.
BESIII has systematically investigated the production and decays of the $\Lcp$~\cite{BESIII:2020nme} for the first time using near-threshold data, which guarantee clean background and controllable systematics.
BESIII implemented the absolute measurement of $\BR({\Lcp \rightarrow p K^-\pi^+})$ using the DT technique~\cite{BESIII:2015bjk} for the first time,
where the relative yields of the DT $\Lcp\Lcm$ pairs over the singly-tagged (ST) $\Lcp$ is counted. The BF is determined as $\BR({\Lcp \rightarrow p K^-\pi^+})=(5.84\pm0.27\pm0.23)\%$~\cite{BESIII:2015bjk}.
This has competitive precision to the result ($6.84\pm0.24^{+0.21}_{-0.27}$)\% reported by Belle~\cite{Belle:2013jfq} at nearly the same time and the combined precision of the two measurements  is 5.2\%, a five-fold reduction of the previous uncertainty~\cite{ParticleDataGroup:2014cgo}.
Since  this mode is the golden channel for detecting  $\Lambda_c^+$ baryons in hadron collider experiments, the BESIII result impacts many aspects of  heavy flavor physics.
For instance,  since the  $\Lambda_b^0$ decays primarily to $\Lcp$~\cite{CLEO:2002sga,Rosner:2012gj},  it constrains the measurement of $|V_{cb}|$ via $\Lambda_b^0\to \Lcp \mu^- \nu_\mu$.
Improved measurements of $\Lcp$ hadronic decay rates can be used to constrain charm and bottom quark fragmentation functions by counting inclusive heavy flavor baryons~\cite{LHCb:2011leg}.
In addition, BESIII also reported other eleven absolute BF results of CF decays with improved precisions, including $\BR(\Lcp \rightarrow p \Ks)=(1.51\pm0.08\pm0.03)\%$, as listed in Table~\ref{tab:LcHad_CFBF}.

For the decay $\Lcp \to pK^-\pi^+$, LHCb has conducted a comprehensive study of intermediate resonance states~\cite{LHCb:2022sck,LHCb:2023crj}.
The study identified contributions from several intermediate resonances, including $p\bar{K}^*_0(700)^0$, $p\bar{K}^*_0(892)^0$, $p\bar{K}^*_0(1430)^0$, $\Delta(1232)^{++}K^-$, $\Delta(1600)^{++}K^-$, $\Delta(1700)^{++}K^-$, $\Lambda(1405)^0\pi^+$, $\Lambda(1520)^0\pi^+$, $\Lambda(1600)^0\pi^+$, $\Lambda(1670)^0\pi^+$, $\Lambda(1690)^0\pi^+$  and $\Lambda(2000)^0\pi^+$.
The fractions and phases of these resonances were determined with high precisions based on partial wave analysis (PWA), providing valuable insights into the dynamics of charmed baryon decays.
Belle studied the $p \km$ invariant mass spectrum in $\Lcp \to pK^-\pi^+$ decay~\cite{Belle:2022cbs}, and found the peaking structure at the $\Lambda(1670)$ resonance can not be well described by a Breit-Wigner lineshape of the $\Lambda(1670)$.  One best fit explains the structure as a $\Lambda \eta$ threshold cusp effect. As the fit projection around the $\Lambda(1670)$ resonance in the LHCb PWA study~\cite{LHCb:2022sck} shows slight deviation from the distribution in data, a future verification of the Belle's claim can be further carried out at LHCb.

Along with the measurement of $\BR(\Lcp \to pK^0_S)$, BESIII also performed the first measurement of the $K_L$-involved process $\Lcp \to pK^0_L$, where the $K_L$ is inferred from the missing mass technique~\cite{BESIII:2024sfz}.
The absolute BF is determined as $\mathcal{B}(\Lambda_c^+\to pK_{L}^{0})=(1.67 \pm 0.06 \pm 0. 04)\%$. In addition, multi-body decays involving $K_L^0$ are also studied, such as $\mathcal{B}(\Lambda_c^+\to pK_{L}^{0}\pi^+\pi^-)=(1.69 \pm 0.10 \pm 0.05)\%$ and $\mathcal{B}(\Lambda_c^+\to pK_{L}^{0}\pi^0)=(2.02 \pm 0.13 \pm 0.05)\%$~\cite{BESIII:2024sfz}.
This work highlights the potential of BESIII in studying the charmed baryon decays into $K_L$, which are extremely challenging in the experiments of hadron collider and $B$ factory.
The $K^0_S-K_{L}^{0}$ asymmetries $R(\Lcp, K^0_{S,L}X)\equiv \frac{B(\Lcp \to K_S^0 X) - B(\Lcp \to K_L^0 X) }{B(\Lcp \to K_S^0 X) + B(\Lcp \to K_L^0 X)}$ are firstly determined to be $R(\Lcp, p K^0_{S,L})=-0.025\pm0.031, R(\Lcp, p K^0_{S,L}\pi^+\pi^-)=-0.027\pm0.048$ and  $R(\Lcp, p K^0_{S,L}\pi^0)=-0.015\pm0.046$, by quoting the BFs in Refs.~\cite{BESIII:2024sfz} and \cite{BESIII:2015bjk}, which show no significant non-zero asymmetries.
The measurements of the $K^0_S-K_{L}^{0}$ asymmetries in charmed baryon decays offer the possibility to access the Doubly CS (DCS) processes involving the neutral kaons and provide further constraints on the CF and DCS amplitudes.

Moreover, BESIII observed, for the first time, the decay mode $\Lambda_c^+\to nK^0_{S}\pi^+$~\cite{BESIII:2016yrc,BESIII:2023pia} with a neutron in the final states.
The BF is measured to be $\BR(\Lambda_c^+\to nK^0_{S}\pi^+)=(1.82\pm0.23\pm0.11)\%$~\cite{BESIII:2016yrc} based on 587\,pb$^{-1}$ data at 4.6\,GeV and updated to be $\BR(\Lambda_c^+\to nK^0_{S}\pi^+)=(1.86\pm0.08\pm0.04)\%$~\cite{BESIII:2023pia} in 2024 with more collision data of 4.5\ifb  between 4.6\,GeV and 4.7\,GeV.
A comparison to $\BR(\Lcp\to pK_S^0\pi^+)$~\cite{BESIII:2015bjk} shows that the ratio $\mathcal{B}(\Lcp\to nK_S^0\pi^+)/\mathcal{B}(\Lcp\to pK_S^0\pi^0)=0.88\pm0.05$, which provides test of isospin symmetry and final state interactions.
Based on the above ratio, the strong phase difference of $I^{(0)}$ and $I^{(1)}$ is calculated to be $\cos\delta=-0.26\pm0.03$, which is one useful experimental input for understanding the final state interactions in $\Lcp$ decays.
More recently, the absolute BF of $\Lcp\to nK_S^0\pi^+\pi^0$ is observed with the BF of $(0.85\pm0.13\pm0.03)\%$ with 9.2$\sigma$ at BESIII~\cite{BESIII:2024xgl}.
These analyses being involved with neutron in the final states were carried out using the missing-mass technique to infer the presence of a neutron, which is straightforward according to the kinematic constrains of  pair production in near-threshold data at BESIII.

$\Lambda_c^+$ CF decays into $\Lambda$ or $\Sigma$ particles 
are also extensively studied.
With the DT technique, the absolute BF for the decays of $\Lambda_c^+ \to \Lambda \pi^+$, $\Lambda \pi^+ \pi^0$ and $\Lambda \pi^+\pi^-\pi^+$ are firstly measured at BESIII~\cite{BESIII:2015bjk}.
The first PWA of the charmed baryon hadronic decay $\Lambda_c^+ \to \Lambda \pi^+ \pi^0$ was performed with the ST method.
From the PWA results, the fit fractions (FFs) and the partial wave amplitudes of intermediate resonances can be derived.
In particular, the corresponding decay asymmetry parameters are determined for the first time.
The significant intermediate processes in $\Lambda_c^+ \to \Lambda \pi^+ \pi^0$ consist of $\Sigma(1385)^{0(+)} \pi^{+(0)}$, $\Sigma(1670)^{0(+)} \pi^{+(0)}$, $\Sigma(1750)^{0(+)} \pi^{+(0)}$, $\Lambda \rho(770)^+$ and $\Lambda\,\mathrm{NR}_{1^-}$~\cite{BESIII:2022udq}.
BESIII also performed the PWA of $\Lcp \to \Lambda \pi^+ \eta$ and find an evidence of the pentaquark candidate $\Sigma(1380)^+$ decaying into $\Lambda \pi^+$.
The BFs for $\Lambda_c^+ \to \Lambda a_0(980)^+$, $\Sigma(1385)^+ \eta$ and $\Lambda(1670) \pi^+$ are obtained based on the PWA results.
The result of $\BR(\Lambda_c^+ \to \Sigma(1385)^+ \eta)$ is consistent with the previous BESIII result~\cite{BESIII:2018qyg} within $2\sigma$, but differs from the Belle result~\cite{Belle:2020xku} by more than $3\sigma$.
The obtained $\BR(\Lcp \to \Lambda(1670) \pi^+)$ is consistent with the Belle result~\cite{Belle:2020xku} within $1\sigma$.
The corresponding updated BF for $\Lcp \to \Lambda \pi^+ \eta$ has best precision and is consistent with the previous results from BESIII~\cite{BESIII:2018qyg} and Belle~\cite{Belle:2020xku}.

The absolute BF for the modes $\Lambda_c^+ \to \Sigma^0 \pi^+$, $\Sigma^+ \pi^0$ and $\Sigma^+ \pi^+ \pi^-$ are firstly measured by BESIII~\cite{BESIII:2015bjk}.
The decays of $\Lcp \to \Sigma^+ \eta$ and $\Sigma^+\eta'$ have been studied by BESIII 
using ST method~\cite{BESIII:2018cdl,BESIII:2025vvd} and the latest update on their BFs 
agree well with the Belle result~\cite{Belle:2022bsi}.
On the theoretical side, most of the calculations~\cite{Korner:1992wi,Zenczykowski:1993jm,Uppal:1994pt,Ivanov:1997ra,Sharma:1998rd,Zou:2019kzq,Geng:2019xbo,Zhao:2018mov} fail to conform with the experimental results of both  $\BR(\Lcp \to \Sigma^+\eta)$ and $\BR(\Lcp\to \Sigma^+\eta')$, and only the recent calculation within the framework of the topological diagram approach and the irreducible SU(3) approach~\cite{Cheng:2025oyr} presents a good agreement  with the both experimental results.
In addition, BESIII implemented the first absolute measurements of the BFs for the $\Sigma^-$-involved CF decays $\Lambda_c^+ \to \Sigma^- \pi^+ \pi^+$ and $\Lambda_c^+ \to \Sigma^- \pi^+ \pi^+ \pi^0$ using the missing-mass and DT technique, as $\Sigma^-$ predominately decays into $n\pi^-$, where the decay $\Lambda_c^+ \to \Sigma^- \pi^+ \pi^+ \pi^0$ is observed for the first time~\cite{BESIII:2017rfd}.

In case of the $\Xi$-involved decays, $\Lambda_c^+ \to \Xi^{*0} K^+$ proceeds only via the $W$-exchange diagram, which is a unique process in charmed baryon decays. 
BESIII for the first time measured the absolute BFs for the decays $\Lambda_c^+ \to \Xi^0 K^+$ and $\Lambda_c^+ \to \Xi(1530)^0 K^+$ using DT technique~\cite{BESIII:2018cvs}, which are obtained as $\BR(\Lambda_c^+ \to \Xi^0 K^+)=(5.90\pm0.86\pm0.39)\times 10^{-3}$ and $\BR(\Lambda_c^+ \to \Xi(1530)^0 K^+)=(5.02\pm0.99\pm0.31)\times 10^{-3}$.
The detection efficiencies of these studies are significantly enhanced by missing the $\Xi^0$ particle in the final state reconstruction, and hence, the precisions are largely improved.
The result of $\BR(\Lambda_c^+ \to \Xi^0 K^+)$ show significant deviations from the previously predicted values~\cite{Korner:1992wi,Zenczykowski:1993hw,Ivanov:1997ra,Sharma:1998rd} by at least $2.6\sigma$.
The measured $\BR(\Lambda_c^+ \to \Xi(1530)^0 K^+)$ favors the calculation in Ref.\cite{Korner:1992wi},
while its prediction on $\BR(\Lambda_c^+ \to \Xi^0 K^+)$ has $4\sigma$ discrepancy from BESIII result.
This indicates that experimental results are essential to calibrate the $W$-exchange diagram amplitudes in these theoretical approaches.
The three-body decays of $\Lambda_c^+ \to \Xi^0 K^+ \pi^0$~\cite{BESIII:2023dvx} and $\Lambda_c^+ \to \Xi^0 K_S^0 \pi^+$~\cite{BESIII:2025rda} are studied for the first time at BESIII based on DT method, in which the resonant intermediate process $\Lambda_c^+ \to \Xi(1530)^0 K^+$ is extracted. The derived $\BR(\Lambda_c^+ \to \Xi(1530)^0 K^+)$ is consistent with the previous BESIII result~\cite{BESIII:2018cvs} within $1\sigma$.

\vspace{0.2cm}
$\bullet$ Singly CS decays
\vspace{0.2cm}

Singly CS (SCS) decays occur predominantly via the tree-level 
$c\to d$ transition. Consequently, their BFs are at least one order of magnitude lower than those of CF decays. Although the penguin diagram contributes only with a minor amplitude to SCS decays, its interference with the tree amplitude can generate CP-violating effects within SM. Therefore, studies of SCS decays provide an important portal for probing strong interaction dynamics and searching for CP violation.

\begin{figure}[tbp]\centering
  \subfigure[]{
  \includegraphics[width=0.45\linewidth]{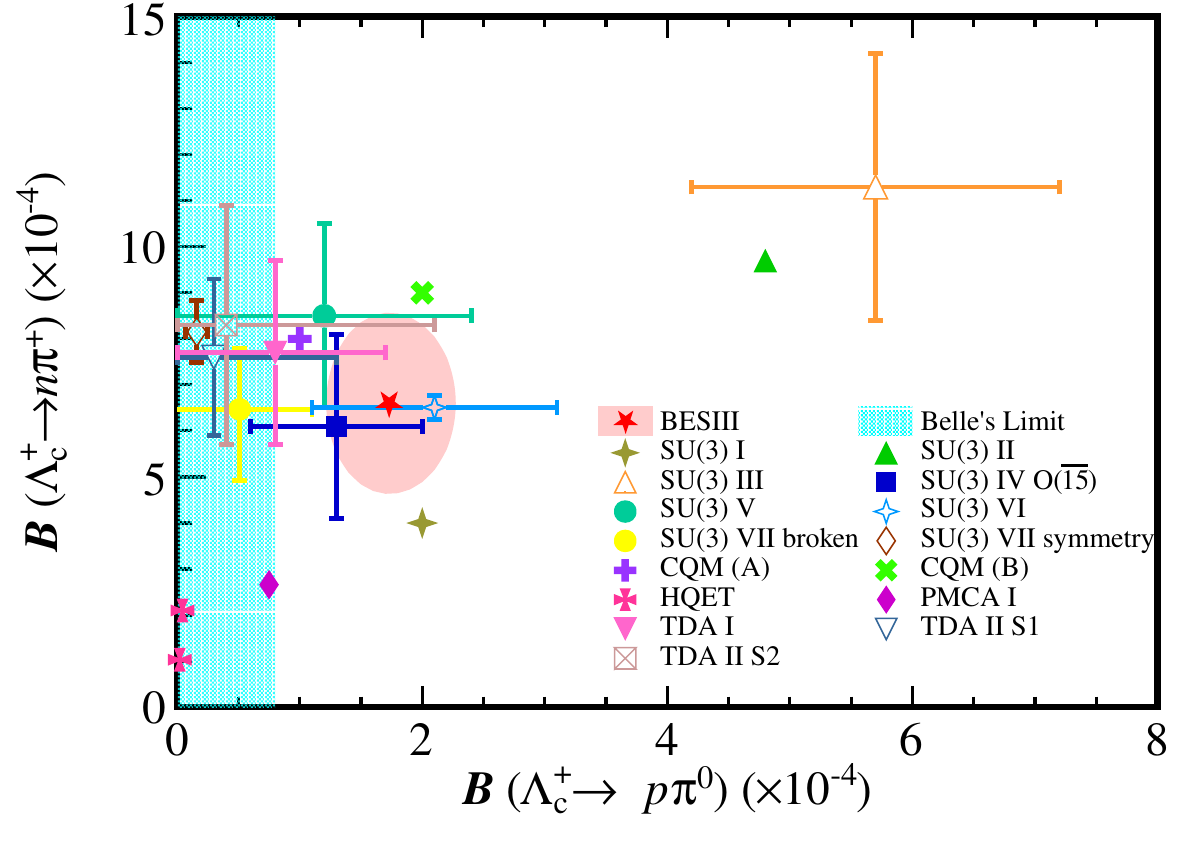}
  \label{fig:ppinpip}
  }
   \subfigure[]{
  \includegraphics[width=0.42\linewidth]{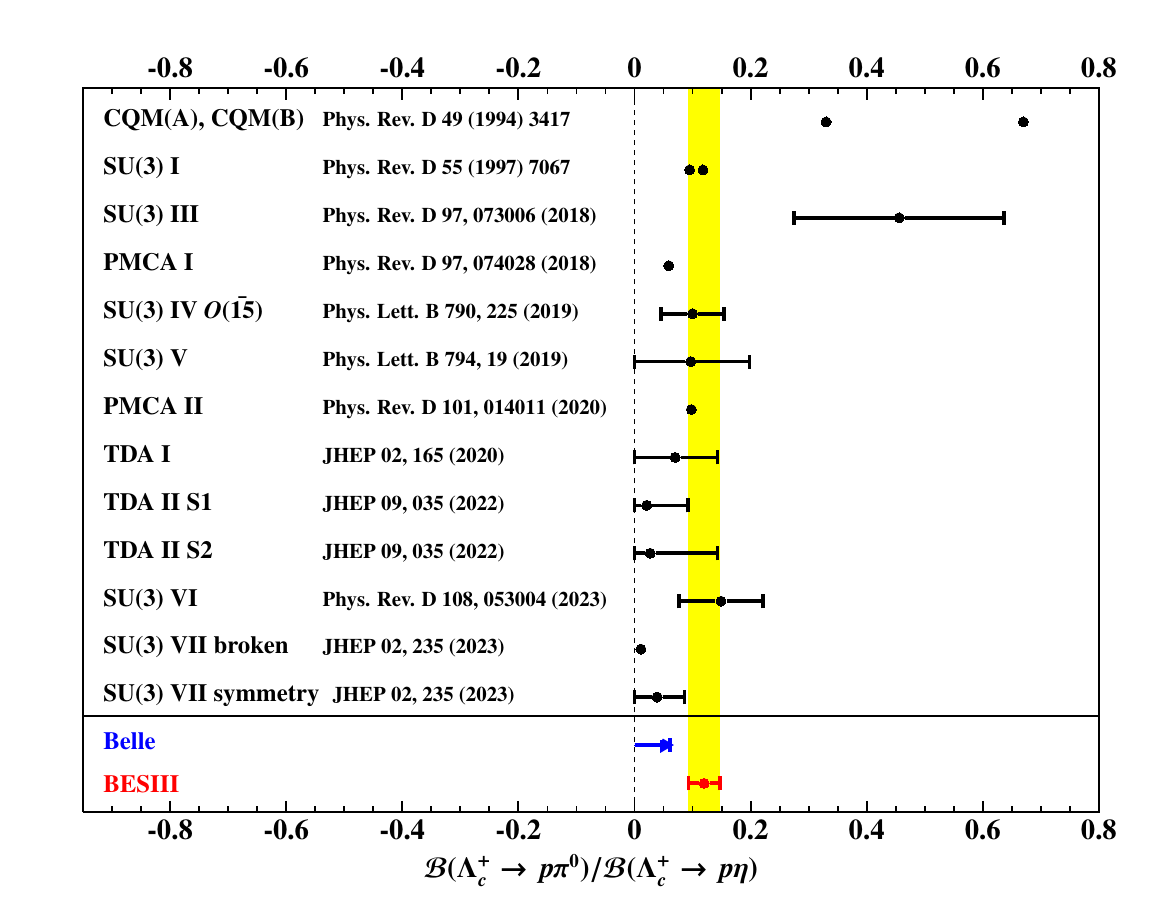}
  \label{fig:ppipeta}
  }
  \caption{Comparison of (a) $\mathcal{B}(\Lcp\to n\pi^+)$ v.s. $\mathcal{B}(\Lcp\to p\pi^0)$  and (b) the ratio $\mathcal{B}(\Lcp\to p\pi^0)/\mathcal{B}(\Lcp\to p\eta )$ with Belle result~\cite{Belle:2021mvw} and previous theoretical calculations piror to the BESIII result of $\mathcal{B}(\Lcp\to p\pi^0)$~\cite{BESIII:2024mgg}. 
The theoretical calculations contain 
constituent quark model (CQM)~\cite{Uppal:1994pt} with two predictions (A) and (B),
heavey quark effective theory (HQET)~\cite{Chen:2002jr}, 
dynamical calculation based on the pole model and current algebra (PMCA)~\cite{Cheng:2018hwl}, topological-diagram approach (TDA) with the results of TDA\,I~\cite{Zhao:2018mov} and TDA\,II~\cite{Hsiao:2021nsc} with solutions S1 and S2, and a few SU(3) flavor symmetry approaches, including SU(3)\,I~\cite{Sharma:1996sc}, 
SU(3)\,II~\cite{Lu:2016ogy}, 
SU(3)\,III~\cite{Geng:2018plk}, 
SU(3)\,IV\,O($\overline{15})$~\cite{Geng:2018rse}, 
SU(3)\,V~\cite{Geng:2019xbo}, 
SU(3)\,VI~\cite{Xing:2023dni} and 
SU(3)\,VII~\cite{Zhong:2022exp} under SU(3) broken and symmetric approaches.
}
\end{figure}

For the two-dody SCS decay with proton final state accompanied by a neutral pseudoscalar meson, there have been  significant progress experimentally, such as the BF measurements of $\Lcp \to p\pi^0$, $p\eta$, $p\eta'$ and $p\omega$ predominantly conducted by BESIII and Belle.
Experimental efforts for $\Lcp\to p\pi^0$ have evolved from upper limits at BESIII~\cite{BESIII:2017fim} and Belle~\cite{Belle:2021mvw} to the first evidence~\cite{BESIII:2023uvs} at BESIII.
Hence, there is no observation reported before 2024.
As the SCS decay rate of $\Lcp\to p\pi^0$ is relatively low, the ST method is useful to sustain high efficiency. However, this induces a high background level, which is a significant challenge for the signal search.
To address the trade-off between signal efficiency and background level, a deep neural network (DNN) is resorted to, 
which has exhibited remarkable capabilities for uncovering new relations and hidden patterns.
Compared to selection-based methods, the topological characteristic of $e^+e^-$ annihilation events can be efficiently recognized and interpreted by a trained DNN model.
This approach culminated in the first observation of $\Lcp\to p\pi^0$ with a statistical significance of 5.4$\sigma$~\cite{BESIII:2024cbr}. The determined BF is $\BR(\Lcp\to p\pi^0)=(1.79\pm0.39\pm0.11\pm0.08)\times10^{-4}$.
This result agrees with the previous BESIII measurement~\cite{BESIII:2017fim,BESIII:2023uvs} and exceeds the Belle upper limit~\cite{Belle:2021mvw} by 2.4$\sigma$.
Figure~\ref{fig:ppinpip} compares the average BESIII $\BR(\Lcp\to p\pi^0)$ result with previous theoretical predictions and Belle measurement.

BESIII conducted the first measurement of the absolute BF as $\mathcal{B}(\Lcp\to p\eta)= (1.24\pm0.28\pm0.10)\times10^{-3}$ with a significance of 4.2$\sigma$~\cite{BESIII:2017fim}.
Subsequently, Belle confirmed the BF with $\mathcal{B}(\Lcp\to p\eta)=(1.42\pm0.05\pm0.11)\times 10^{-3}$~\cite{Belle:2021mvw}.
With increased data statistics at BESIII, the BF is updated to be $\mathcal{B}(\Lcp\to p\eta)=(1.57\pm0.11\pm0.04)\times10^{-3}$ ~\cite{BESIII:2023ooh} and $\mathcal{B}(\Lcp\to p\eta)=(1.63\pm0.31\pm0.11)\times10^{-3}$~\cite{BESIII:2023uvs} based on ST and DT methods, respectively. 
Through a rare decay channel $\eta\to \mu^+\mu^-$, LHCb obtained the BF as $\mathcal{B}(\Lcp\to p\eta)=(1.67\pm0.69\pm0.23\pm0.34)\times10^{-3}$~\cite{LHCb:2024hju}.
$\Lcp\to p\eta'$ is observed for the first time at Belle with $\mathcal{B}(\Lcp\to p\eta')=(4.73\pm0.82\pm0.46\pm0.24)\times 10^{-4}$~\cite{Belle:2021vyq}. Later, BESIII reported the absolute BF for $\Lcp\to p\eta'$ to be $(5.62^{+2.46}_{-2.04}\pm0.26)\times10^{-4}$ \cite{BESIII:2022izy}, which is consistent with the Belle result within uncertaintie.

In the anslysis of $\Lcp\to p \mumu$ using RUN1 data at LHCb, a significant $\Lcp\to p\omega$ signal was observed in the dimuon mass spectrum with $\BR(\Lcp\to p \omega) =(9.4 \pm 3.2\pm 1.0\pm 2.0)\times 10^{-4}$\cite{LHCb:2017yqf}, where the third uncertainty is from the limited knowledge of the BF of the reference mode $\Lcp \to p \phi, \phi \to \mumu$. This result was later updated to be $\mathcal{B}(\Lcp\to p \omega)=(9.82\pm1.23\pm0.73\pm2.79)\times10^{-4}$ in the same process based on RUN2 dataset at LHCb~\cite{LHCb:2024hju}.
Afterward, Belle improved the precision of the BF as $\mathcal{B}(\Lcp\to p \omega)=(8.27\pm0.75\pm0.62\pm0.42)\times 10^{-4}$~\cite{Belle:2021btl}.
Most recently, BESIII reported the most precise measurement as $\mathcal{B}(\Lcp\to p \omega)=(1.11\pm0.20\pm0.07)\times10^{-3}$~\cite{BESIII:2023ooh} based on the ST method.
In addition, the BF of $\Lcp\to p \rho$ is reported to be $\mathcal{B}(\Lcp\to p \rho)=(1.52\pm0.34\pm0.14\pm0.24)\times10^{-3}$~\cite{LHCb:2024hju} in the process $\Lcp\to p \mumu$ at LHCb. This result can be cross-checked through the analysis of intermediate $\rho$ contributions in $\Lcp\to p \pip\pim$ in the future.

The SCS decay involving a neutron in the final states, $\Lcp\to n\pi^+$, is observed for the first time with a statistical significance of 7.3$\sigma$ at BESIII~\cite{BESIII:2022bkj} and the BF is measured to be $\BR(\Lcp\to n\pi^+) = (6.6\pm1.2\pm0.4)\times10^{-4}$.
When quoting the result of $\BR(\Lcp\to p\pi^0)$, the ratio $\BR(\Lcp\to n\pi^+)/\BR(\Lcp\to p\pi^0)$ is calculated as $3.7\pm1.1$.
As the comparison shown in Fig.~\ref{fig:ppinpip}, the BFs for $\BR(\Lcp\to n\pi^+$ and $\BR(\Lcp\to p\pi^0)$ provide crucial tests on a avarity of theorectical calculations in charmed baryon decays.
The ratio $\BR(\Lcp\to p\pi^0)/\BR(\Lcp\to p\eta)$ is derived as $(12.0\pm2.6\pm0.7)\%$~\cite{BESIII:2024cbr}, much less that one, because $\Lcp\to p\eta$($\Lcp\to p\pi^0$) has a large constructive(destructive) interference between the factorizable and nonfactorizable amplitudes for both $S$ and $P$ waves~\cite{Cheng:2018hwl}.
This experimetnal ratio is compared to different theorectical calculaitons, as shown in Fig.~\ref{fig:ppipeta}.
results resolve the longstanding discrepancy between earlier experimental searches, providing both a decisive conclusion and valuable input for QCD-inspired theoretical models.

The three-body SCS decay $\Lcp\to p\pip\pim$ and $\Lcp \to p\kp\km$ have been measured with high precision in the BESIII and LHCb experiments.
BESIII reported $\BR(\Lcp\to p \pip\pim)=(3.91\pm0.28\pm0.15\pm0.24)\times10^{-3}$ and $\BR(\Lcp\to p (\kp\km)_{\mathrm{non-}\phi})=(5.47\pm1.30\pm0.41\pm0.33)\times10^{-4}$ together with the $\phi$ contribution of $\BR(\Lcp\to p \phi)=(1.06\pm0.19\pm0.08\pm0.06)\times10^{-3}$~\cite{BESIII:2016ozn}.
LHCb has also measured the BFs for $\Lcp\to p\kp\km$ and $\Lcp \to p\pi^+\pi^-$, with results of $\BR(\Lcp\to p \pip\pim)=(4.72 \pm 0.05 \pm 0.11 \pm 0.25)\times10^{-3}$ and $\BR(\Lcp\to p \kp\km)=(1.08 \pm 0.02 \pm 0.02 \pm 0.06)\times10^{-3}$~\cite{LHCb:2022sck}, where the latter includes the both $\phi$ contributio and the non-$\phi$ contributions taking the input BF for the reference mode $\BR(\Lcp\to p\km\pip)=(6.35 \pm 0.33)\%$ from the 2016 version of PDG~\cite{ParticleDataGroup:2016lqr}.
These results for the SCS decay modes are consistent within the uncertainties.
In addition, BESIII reported the first evidence for the SCS decay $\Lcp\to nK_S^0K^+$~\cite{BESIII:2023pia} and observation of the SCS decay $\Lcp\to n \pi^+\pi^0$~\cite{BESIII:2022xne}, where the neutron signls are obtained by the missing-mass technique based on the DT method.

The BF asymmetry between the two charge-conjugate modes $A^{\Lcp}_{CP}(f)=\frac{\BR(\Lcp \to f)-\BR(\Lcm \to \bar{f})}{\BR(\Lcp \to f)+\BR(\Lcm \to \bar{f})}$ can be used to test direct CP violation, where $f$ denotes decay final states.
The first attempt of searching for CP violation in $\Lcp$ decays is constituted with  in the SCS decays $\Lcp\to p\kp\km$ and $\Lcp \to p\pi^+\pi^-$ by LHCb.
To cancel the production and detection asymmetries, the difference of the CP asymmetries is measured to be $\Delta A^{\Lcp}_{CP}=A^{\Lcp}_{CP}(p\kp\km)-A^{\Lcp}_{CP}(p\pi^+\pi^-)=(0.31\pm0.91\pm0.61)\%$, which is consistent with zero asymmetry. 
Belle II recently measured separate CP asymmetries $A^{\Lcp}_{CP}(p\kp\km)=(3.9\pm1.7\pm0.7)\%$ and $A^{\Lcp}_{CP}(p\pi^+\pi^-)=(0.3\pm1.0\pm0.2)\%$~\cite{Belle-II:2025xvc}, which agree with CP conservation.
The CP asymmetry in $\Lcp\to p \mu^+\mu^-$ decays around the $\phi$ resonance, which is dominated by the long-distance $\phi$ contributions, are investigated at LHCb~\cite{LHCb:2025bfy}, which gives $A^{\Lcp}_{CP}(p (\mu^+\mu^-)_{\phi})=(-1.1\pm4.0\pm0.5)$\%.
In the dimoun mass region around the $\phi$ resonance, the CP average ($\Sigma A_{FB}$) and CP asymmetry ($\Delta A_{\rm FB}$) of the forward-backward asymmetry in the muon system of $\Lcp\to p\mu^+\mu^-$ decays is reported as
$\Sigma A_{FB}=(3.9\pm4.0\pm0.6)\%$ and $\Delta A_{FB}=(3.1\pm4.0\pm0.4)\%$~\cite{LHCb:2025bfy}.
These results are consistent with the conservation of CP symmetry and the SM expectations.
Further more data statistics is required to match the sensitivity of CP asymmetry measurements in charmed mesons.

For the $\Lambda_c^+$ CS decays involving hyperon states, only $\Lambda$ or $\Sigma$ can be produced.  as the sum of the $\Xi \overline{K} \overline{K}$ particles exceeds the $\Lcp$ mass.
In these aspects, there have been good experimental progress from BESIII and Belle.
The BF for $\Lambda_c^+ \to \Lambda K^+$ was measured to be $\BR(\Lambda_c^+ \to \Lambda K^+)=(6.21\pm0.44\pm0.26\pm0.34)\times 10^{-4}$ by BESIII using the ST method~\cite{BESIII:2022tnm}, which is consistent with the Belle result $(6.57\pm0.17\pm0.11\pm0.35)\times 10^{-4}$~\cite{Belle:2022uod}. These results disagree with some theorectical calculations, such as constituent quark model~\cite{Uppal:1994pt} and current algebra~\cite{Cheng:2018hwl}.
The decays $\Lambda_c^+ \to \Sigma^0 K^+$ and $\Lambda_c^+ \to \Sigma^+ K_S^0$ were studied by BESIII using the ST method~\cite{BESIII:2022wxj}, where $
\BR(\Lambda_c^+ \to \Sigma^+ K_S^0)$ decay was measured for the first time.
Later Belle improved the result of $\BR(\Lambda_c^+\to\Sigma^0 K^+)$ with better precision~\cite{Belle:2022uod}.
The ratio $\frac{\BR(\Lambda_c^+\to\Sigma^0 K^+)}{\BR(\Lambda_c^+\to\Sigma^+K_S^0)}$ is found to be consistent with the predictions in Refs.~\cite{Zou:2019kzq,Geng:2019xbo} under SU(3) flavor symmetry, while the experimental result of $\BR(\Lambda_c^+\to\Sigma^+K_S^0)$ is generally compatible with the predictions in Refs.~\cite{Uppal:1994pt,Zou:2019kzq,Zhao:2018mov,Geng:2019xbo} within $1\sim2\sigma$.
For decays $\Lambda_c^+ \to \Lambda K^+$ and $\Lambda_c^+ \to \Sigma^0 K^+$, the first searches for direct CP asymmetry in two-body SCS decays of charmed baryons are implemented by Belle~\cite{Belle:2022uod}, which gives $A^{\Lcp}_{CP}(\Lambda K^+)=0.021\pm0.026\pm0.001$ and $A^{\Lcp}_{CP}(\Sigma^0 K^+)=0.025\pm0.054\pm0.004$.
These are consistent with  CP symmetry in charmed baryon decays.
For multi-body decays, BESIII observed a few three-body decays, such as $\Lambda_c^+ \to \Lambda K_S^0 \pi^+$~\cite{BESIII:2024xny}, $\Lambda_c^+ \to \Lambda K^+ \pi^0$~\cite{BESIII:2023sdr} and $\Lambda_c^+ \to \Sigma^- K^+ \pi^+$~\cite{BESIII:2023iwu},  and found evidence of the four-body decay $\Lambda_c^+ \to \Lambda K^+ \pi^+ \pi^-$~\cite{BESIII:2023sdr}.

\vspace{0.2cm}
$\bullet$ Doubly CS decays
\vspace{0.2cm}

While several DCS decays of charmed mesons have been observed, DCS decays of charmed baryons had not
yet been observed before 2015, due to the smaller production cross sections for charmed
baryons in experiment.
In 2015, Belle reported the first observation of the decay $\Lcp \to pK^+\pi^-$  using a 980\,\ifb data sample. This 
is the first and the only DCS decay observed in experiment inside the charmed baryon sector.
The relative BF is measured to be $\BR(\Lcp\to p K^+ \pi^−)/\BR(\Lcp\to p K^- \pi^+)=(2.35\pm 0.27 \pm 0.21)\times 10^{−3}$, which correspond to $\BR(\Lcp\to p K^+ \pi^−)=(1.61\pm0.23^{+0.07}_{-0.08})\times10^{-4}$~\cite{Belle:2015wxn}.
Later on, LHCb improved the BF ratio as $\BR(\Lcp\to p K^+ \pi^−)/\BR(\Lcp\to p K^- \pi^+)=(1.65\pm 0.15 \pm 0.05)\times 10^{−3}$~\cite{LHCb:2017xtf}, which is lower than the Belle value by $2.0\sigma$.
The obtained BF at LHCb is given as  $\BR(\Lcp \to pK^+\pi^-)=(1.04 \pm 0.09 \pm 0.03 \pm 0.05)\times10^{-4}$~\cite{LHCb:2017xtf}.
The ratio  $\BR(\Lcp\to p K^+ \pi^−)/\BR(\Lcp\to p K^- \pi^+)$ is a useful variable with which to indirectly study the role of $W$-exchange process in charmed baryon hadronic decays. In the absence of SU(3) flavour symmetry
breaking, the ratio can naively be expected to be equal to $\tan^4 \theta_c$, where $\theta_c$ is the
Cabibbo mixing angle. 
As there are a fuitful resonant contributions in the reference decay mode of  $\Lcp\to p K^- \pi^+$, detailed amplitue analysis on the DCS decay $\Lcp\to p K^+ \pi^−$ would be important to disentangle the intermediate resonance contributions, which can provide more direct information on the $W$-exchange process and $SU(3)$ symmetry in charmed baryon decays.

\vspace{0.2cm}
$\bullet$ Inclusive decays
\vspace{0.2cm}

The BF for the inclusive hadrnoic decays provide overall constrains on different types of exculsive decay rates.
The inclusive BF of both $\Lcp\to p X$ and $\Lcp\to n X$ were estimated to be $(50\pm16)\%$, inferred from the known exclusive $B$-meson decays and the fact that all $\Lcp$ particles must decay into either proton or neutron~\cite{CLEO:1991ejj}.
The  experimental determination on the inclusive BF provides direct test on whether there exists a significant difference between the decays of $\Lcp$ with a proton and a neutron in the final states.
Based on the DT method, BESIII meausred the absolute BF of the inclusive decay $\Lcm \to \bar{n} X$ as $\mathcal{B}(\bar{\Lambda}_c^-\to \bar{n}X) = (32.4\pm0.7\pm1.5)\%$, where $X$ refers to any possible particle system~\cite{BESIII:2022onh}.
Presently, the sum of experimentally measured exclusive decay rates with a neutron(antineutron) in the final states is $(26.54\pm0.72)\%$~\cite{ParticleDataGroup:2024cfk}.
Assuming CP symmetry, the BESIII result indicates that about $5.86\%$ of $\Lcp(\bar{\Lambda}_c^-)$ decay modes with a neutron(antineutron) involved have not been observed.

The $\Lcp$ inclusive decays into $\Lambda$ are mostly governed by the $c-s$ transition, which is a prominent process in the charmed baryon decays.
Hence, the BF of $\Lcp\to \Lambda X$ provides essential input in the calculation of the lifetimes of charmed baryons, as current theoretical treatement suffer from large uncertainties. Furthermore, decay dynamics in the $\Lcp\to \Lambda X$ would benefit the research on heavier charmed baryons.
With the DT technique, The absolute BF of $\Lcp\to \Lambda X$ is measured to be $\mathcal{B}(\Lambda_c^+\to \Lambda X) = (38.2^{+2.8}_{-2.2}\pm0.9)\%$ by BESIII~\cite{BESIII:2018ciw}.
The sum of experimentally known exclusive decay rates involved with a $\Lambda$ is $(31.98\pm1.20)\%$~\cite{ParticleDataGroup:2024cfk}.
Therefore, there are still space to explore more decay modes consisting of a $\Lambda$.
In addition, the direct CP violation in the charge asymmetry of this inclusive decay is obtained as $A^{\Lcp}_{CP}(\Lambda X)=(2.1^{+7.0}_{-6.6}\pm1.6)\%$, in which no CP violation is observed.

In addition, BESIII determined the absolute BF of the inclusive $K_S^0$ decays $\Lcp\to K_S^0 X$ to be $\mathcal{B}(\Lambda_c^+\to K_S^0 X) = (10.9\pm0.2 \pm0.1)\%$~\cite{BESIII:2020cpu,BESIII:2025cel}.
Summing over the known BFs for the final states containing $K_S^0$ gives a rate of $(8.77\pm0.78)\%$~\cite{ParticleDataGroup:2024cfk}. So there remains about $2\%$ rate of unknown decays involving $K_S^0$.


\begin{table*}[tbp]
  \caption{ Measurements of the BFs for the CF decays of the $\Lcp$~(in units of $\%$).}
  \vspace{-0.5cm}
  \label{tab:LcHad_CFBF}
  \begin{center}
    \footnotesize
    \begin{tabular}{lcc|lcc}
      \hline\hline
      Mode                                                           & BF            & Experiment                        & Mode                                        & BF                     & Experiment                             \\
      \hline\hline
      \multicolumn{6}{l}{\textbf{Nucleon-involved}}                                                                                                                                                                                      \\ \hline
      $\Lcp\to pK^0_S$                                               & $1.52\pm0.09$ & BESIII(2016)\cite{BESIII:2015bjk} & \multirow{2}*{$\Lcp\to n K^0_S\pi^+$}       & $1.82\pm0.25$          & BESIII(2017)\cite{BESIII:2016yrc}      \\
      \cline{1-3}
      $\Lcp\to pK^0_L$                                               & $1.67\pm0.07$ & BESIII(2024)\cite{BESIII:2024sfz} & ~                                           & $1.86\pm0.09$          & BESIII(2024)\cite{BESIII:2023pia}      \\
      \hline
      $\Lcp\to p\bar{K}^*_0(700)^0\to pK^-\pi^+$                     & $0.19\pm0.06$ & LHCb(2023)\cite{LHCb:2022sck}     & $\Lcp\to n K^0_S \pi^+\pi^0$                & $0.85\pm0.13$          & BESIII(2024)\cite{BESIII:2024xgl}      \\
      \hline
      $\Lcp\to p\bar{K}^*_0(892)^0\to pK^-\pi^+$                     & $1.38\pm0.08$ & LHCb(2023)\cite{LHCb:2022sck}     & $\Lcp\to n K^- \pi^+\pi^+$                  & $1.90\pm0.12$          & BESIII(2023)\cite{BESIII:2022xne}      \\
      \hline
      $\Lcp\to p\bar{K}^*_0(1430)^0\to pK^-\pi^+$                    & $0.92\pm0.18$ & LHCb(2023)\cite{LHCb:2022sck}     & \multirow{2}*{$\Lcp\to pK^0_S\pi^0$}        & $1.87\pm0.14$          & BESIII(2016)\cite{BESIII:2015bjk}      \\
      \cline{1-3}
      $\Lcp\to \Delta(1232)^{++}K^-\to p\pi^+K^-$                    & $1.78\pm0.05$ & LHCb(2023)\cite{LHCb:2022sck}     & ~                                           & $2.12\pm0.11$          & Belle(II)(2025)\cite{Belle-II:2025bgw} \\
      \hline
      $\Lcp\to \Delta(1600)^{++}K^-\to p\pi^+K^-$                    & $0.28\pm0.10$ & LHCb(2023)\cite{LHCb:2022sck}     & $\Lcp\to pK^0_L\pi^0$                       & $2.02\pm0.14$          & BESIII(2024)\cite{BESIII:2024sfz}      \\
      \hline
      $\Lcp\to \Delta(1700)^{++}K^-\to p\pi^+K^-$                    & $0.24\pm0.06$ & LHCb(2023)\cite{LHCb:2022sck}     & \multirow{2}*{$\Lcp\to pK^0_S\eta$}         & $0.41\pm0.09$          & BESIII(2021)\cite{BESIII:2020kzc}      \\
      \cline{1-3}
                                                                     &               &                                   & ~                                           & $0.44\pm0.03$          & Belle(2023)\cite{Belle:2022pwd}        \\
      \cline{4-6}
                                                                     &               &                                   & $\Lcp\to pK^0_S\pi^+\pi^-$                  & $1.53\pm0.14$          & BESIII(2016)\cite{BESIII:2015bjk}      \\
      \cline{4-6}
                                                                     &               &                                   & $\Lcp\to pK^0_L\pi^+\pi^-$                  & $1.69\pm0.11$          & BESIII(2024)\cite{BESIII:2024sfz}      \\
      \cline{4-6}
                                                                     &               &                                   & \multirow{2}*{$\Lcp\to pK^-\pi^+$}          & $6.84^{+0.32}_{-0.36}$ & Belle(2014)\cite{Belle:2013jfq}        \\
                                                                     &               &                                   & ~                                           & $5.84\pm0.35$          & BESIII(2016)\cite{BESIII:2015bjk}      \\
      \cline{4-6}
                                                                     &               &                                   & \multirow{2}*{$\Lcp\to pK^-\pi^+\pi^0$}     & $4.53\pm0.38$          & BESIII(2016)\cite{BESIII:2015bjk}      \\
                                                                     &               &                                   & ~                                           & $4.42\pm0.21$          & Belle(2017)\cite{Belle:2017tfw}        \\
      \hline\hline
      \multicolumn{6}{l}{\textbf{$\Lambda$-involved}}                                                                                                                                                                     \\ \hline
      \multirow{2}*{$\Lcp\to \Lambda\pi^+$}                          & $1.24\pm0.08$ & BESIII(2016)\cite{BESIII:2015bjk} & $\Lcp\to \Lambda\pi^+\pi^0$                 & $7.01\pm0.42$          & BESIII(2016)\cite{BESIII:2015bjk}      \\
      \cline{4-6}
      ~                                                              & $1.31\pm0.09$ & BESIII(2023)\cite{BESIII:2022bkj} & \multirow{3}*{$\Lcp\to \Lambda\pi^+\eta$}   & $1.84\pm0.26$          & BESIII(2019)\cite{BESIII:2018qyg}      \\
      \cline{1-3}
      $\Lcp\to \Lambda\rho(770)^+$                                   & $4.06\pm0.52$ & BESIII(2022)\cite{BESIII:2022udq} & ~                                           & $1.84\pm0.13$          & Belle(2021)\cite{Belle:2020xku}        \\
      \cline{1-3}
      $\Lcp\to \Lambda a_0(980)^+$                                   & $1.23\pm0.21$ & BESIII(2025)\cite{BESIII:2018qyg} & ~                                           & $1.94\pm0.13$          & BESIII(2025)\cite{BESIII:2024mbf}      \\
      \hline
      $\Lcp\to \Lambda(1405)\pi^+\to pK^-\pi^+$                      & $0.48\pm0.19$ & LHCb(2023)\cite{LHCb:2022sck}     & $\Lcp\to \Lambda\pi^+\pi^-\pi^+$            & $3.81\pm0.30$          & BESIII(2016)\cite{BESIII:2015bjk}      \\
      \hline
      $\Lcp\to \Lambda(1520)\pi^+\to pK^-\pi^+$                      & $0.12\pm0.02$ & LHCb(2023)\cite{LHCb:2022sck}     & \multirow{2}*{$\Lcp\to \Lambda K^0_S K^+$}                 & $0.30\pm0.03$          & BESIII(2025)\cite{BESIII:2024xny}      \\
      \cline{1-3}
      $\Lcp\to \Lambda(1600)\pi^+\to pK^-\pi^+$                      & $0.32\pm0.12$ & LHCb(2023)\cite{LHCb:2022sck}     &  &  $0.31\pm0.05$          & BESIII(2025)\cite{BESIII:2025rda}  
      \\  \hline
      $\Lcp\to \Lambda(1670)\pi^+\to pK^-\pi^+$                      & $0.07\pm0.02$ & LHCb(2023)\cite{LHCb:2022sck}     &                                             &                        &                                        \\
      \cline{1-3}
      \multirow{2}*{$\Lcp\to \Lambda(1670)\pi^+\to\Lambda\eta\pi^+$} & $0.27\pm0.06$ & Belle(2021)\cite{Belle:2020xku}   &                                             &                        &                                        \\
      ~                                                              & $0.27\pm0.06$ & BESIII(2025)\cite{BESIII:2024mbf} &                                             &                        &                                        \\
      \cline{1-3}
      $\Lcp\to \Lambda(1690)\pi^+\to pK^-\pi^+$                      & $0.07\pm0.02$ & LHCb(2023)\cite{LHCb:2022sck}     &                                             &                        &                                        \\
      \cline{1-3}
      $\Lcp\to \Lambda(2000)\pi^+\to pK^-\pi^+$                      & $0.60\pm0.07$ & LHCb(2023)\cite{LHCb:2022sck}     &                                             &                        &                                        \\
      \hline\hline
      \multicolumn{6}{l}{\textbf{$\Sigma$-involved}}                                                                                                                                                                      \\ \hline
      $\Lcp\to \Sigma^+\pi^0$                                        & $1.18\pm0.10$ & BESIII(2016)\cite{BESIII:2015bjk} & \multirow{2}*{$\Lcp\to \Sigma^+\pi^+\pi^-$} & $4.25\pm0.31$          & BESIII(2016)\cite{BESIII:2015bjk}      \\
      \cline{1-3}
      \multirow{3}*{$\Lcp\to \Sigma^+\eta$}                          & $0.41\pm0.20$ & BESIII(2018)\cite{BESIII:2018cdl} & ~                                           & $4.57\pm0.28$          & Belle(2018)\cite{Belle:2018gcs}        \\
      \cline{4-6}
      ~                                                              & $0.31\pm0.05$ & Belle(2023)\cite{Belle:2022bsi}   & $\Lcp\to \Sigma^+\pi^0\pi^0$                & $1.57\pm0.15$          & Belle(2018)\cite{Belle:2018gcs}        \\
    \cline{4-6}
      
      & $0.38\pm0.06$ & BESIII(2025)\cite{BESIII:2025vvd} & $\Lcp\to \Sigma^0\pi^+\pi^0$                & $3.65\pm0.30$          & Belle(2018)\cite{Belle:2018gcs}        \\
      \hline

      \multirow{3}*{$\Lcp\to \Sigma^+\eta'$}                         & $1.34\pm0.56$ & BESIII(2018)\cite{BESIII:2018cdl}  & $\Lcp\to \Sigma^0\pi^+\eta$                 & $0.76\pm0.08$          & Belle(2021)\cite{Belle:2020xku}       \\
      \cline{4-6}
      ~                                                              & $0.42\pm0.09$ & Belle(2023)\cite{Belle:2022bsi}  & $\Lcp\to \Sigma^-\pi^+\pi^+$                & $1.81\pm0.19$          & BESIII(2017)\cite{BESIII:2017rfd}  
      \\     \cline{4-6}

      & $0.57\pm0.18$ & BESIII(2025)\cite{BESIII:2025vvd}
      & $\Lcp\to \Sigma^-\pi^+\pi^+\pi^0$           & $2.11\pm0.36$          & BESIII(2017)\cite{BESIII:2017rfd}   
        \\ \hline
      
      $\Lcp\to \Sigma^+\omega$                                       & $1.56\pm0.21$ & BESIII(2016)\cite{BESIII:2015bjk} 
      & $\Lcp\to\Sigma^+K^+K^-$                     & $0.38\pm0.05$          & BESIII(2023)\cite{BESIII:2023rky}    
      \\ \hline

      $\Lcp\to \Sigma^+\phi$                                         & $0.41\pm0.09$ & BESIII(2023)\cite{BESIII:2023rky} 
      & $\Lcp\to\Sigma^+K^+K^-_{\text{non-}\phi}$   & $0.20\pm0.04$          & BESIII(2023)\cite{BESIII:2023rky}    
      \\ \hline
      \multirow{2}*{$\Lcp\to \Sigma^0\pi^+$}                         & $1.27\pm0.09$ & BESIII(2016)\cite{BESIII:2015bjk}   
        & $\Lcp\to \Sigma^0 K^0_S K^+$ &  $0.08\pm0.03$          & BESIII(2025)\cite{BESIII:2025rda}                                  \\ \cline{4-6}
      ~                                                              & $1.22\pm0.11$ & BESIII(2023)\cite{BESIII:2022bkj} &                                             &                        &                                       \\
      \cline{1-3}

      $\Lcp\to \Sigma(1385)^+\pi^0$                                  & $0.59\pm0.08$ & BESIII(2022)\cite{BESIII:2022udq} &                                             &                        &                                        \\
      \cline{1-3}
      \multirow{3}*{$\Lcp\to \Sigma(1385)^+\eta$}                    & $0.91\pm0.20$ & BESIII(2019)\cite{BESIII:2018qyg} &                                             &                        &                                        \\
      ~                                                              & $1.21\pm0.12$ & Belle(2021)\cite{Belle:2020xku}   &                                             &                        &                                        \\
      ~                                                              & $0.68\pm0.08$ & BESIII(2025)\cite{BESIII:2024mbf} &                                             &                        &                                        \\
      \cline{1-3}
      $\Lcp\to \Sigma(1385)^0\pi^+$                                  & $0.65\pm0.10$ & BESIII(2022)\cite{BESIII:2022udq} &                                             &                        &                                        \\
      \hline\hline
      \multicolumn{6}{l}{\textbf{$\Xi$-involved}}                                                                                                                                                                         \\ \hline
      $\Lcp\to \Xi^0K^+$                                             & $0.59\pm0.09$ & BESIII(2018)\cite{BESIII:2018cvs} & $\Lcp\to\Xi^0K^+\pi^0$                      & $0.78\pm0.17$          & BESIII(2024)\cite{BESIII:2023dvx}      \\ \hline
 
      \multirow{2}*{$\Lcp\to \Xi(1530)^0K^+$}                        & $0.50\pm0.10$ & BESIII(2018)\cite{BESIII:2018cvs} 
      & $\Lcp\to \Xi^0 K^0_S \pi^+$ &  $0.37\pm0.06$          & BESIII(2025)\cite{BESIII:2025rda}   \\ \cline{4-6}
      ~                                                              & $0.60\pm0.11$ & BESIII(2024)\cite{BESIII:2023dvx} &                                             &                        &                                        \\
      \hline \hline
    \end{tabular}
  \end{center}
\end{table*}

\begin{table*}[tbp]
  \caption{The determined BFs for the CS decays of the $\Lcp$~(in units of $10^{-3}$). Upper limits are set at 90\% confidence level.} \label{tab:LcHad_CSBF}
  \begin{center}
    \footnotesize
    \begin{tabular}{lcc|lcc}
      \hline\hline
      Mode                                    & BF                                  & Experiment                        & Mode                                     & BF                     & Experiment                        \\
      \hline\hline
      \multicolumn{6}{l}{\textbf{Nucleon-involved}}                                                                                                                                                                             \\ \hline
      $\Lcp\to n\pi^+$                        & $0.66\pm0.13$                       & BESIII(2022)\cite{BESIII:2022bkj} & $\Lcp\to nK^+\pi^0$                      & $<0.71$                & BESIII(2024)\cite{BESIII:2023dvx} \\
      \hline
      \multirow{4}*{$\Lcp\to p\pi^0$}         & $<0.27$                             & BESIII(2017)\cite{BESIII:2017fim} & $\Lcp\to n \pi^+\pi^0$                   & $0.64\pm0.09$          & BESIII(2023)\cite{BESIII:2022xne} \\
      \cline{4-6}
      ~                                       & $<0.08$                             & Belle(2021)\cite{Belle:2021mvw}   & $\Lcp\to n K^0_S K^+$                    & $0.39^{+0.17}_{-0.14}$ & BESIII(2024)\cite{BESIII:2023pia} \\
      \cline{4-6}
      ~                                       & $0.16^{+0.07}_{-0.06}$              & BESIII(2024)\cite{BESIII:2023uvs} & $\Lcp\to n \pi^+\pi^-\pi^+$              & $0.45\pm0.08$          & BESIII(2023)\cite{BESIII:2022xne} \\
      \cline{4-6}
      ~                                       & $0.18\pm0.04$                       & BESIII(2025)\cite{BESIII:2024cbr} & \multirow{2}*{$\Lcp\to p\pi^+\pi^-$}     & $3.91\pm0.40$          & BESIII(2016)\cite{BESIII:2016ozn} \\
      \cline{1-3}
      \multirow{5}*{$\Lcp\to p\eta$}          & $1.24\pm0.30$                       & BESIII(2017)\cite{BESIII:2017fim} & ~                                        & $4.72\pm0.28$          & LHCb(2018)\cite{LHCb:2017xtf}     \\
      \cline{4-6}
      ~                                       & $1.42\pm0.12$                       & Belle(2021)\cite{Belle:2021mvw}   & $\Lcp\to pK^+K^-$                        & $1.08\pm0.07$          & LHCb(2018)\cite{LHCb:2017xtf}     \\
      \cline{4-6}
      ~                                       & $1.57\pm0.12$                       & BESIII(2023)\cite{BESIII:2023ooh} & $\Lcp\to p(K^+K^-)_{\text{non-}\phi}$      & $0.55\pm0.14$          & BESIII(2016)\cite{BESIII:2016ozn} \\
      \cline{4-6}
      ~                                       & $1.63\pm0.33$                       & BESIII(2024)\cite{BESIII:2023uvs} & $\Lcp\to pK^0_SK^0_S$                    & $0.24\pm0.02$          & Belle(2023)\cite{Belle:2022pwd}   \\
      \cline{4-6}
      ~                                       & $1.67\pm0.80$                       & LHCb(2024)\cite{LHCb:2024hju}     & $\Lcp\to p\phi\pi^0$                     & $<0.15$                & Belle(2017)\cite{Belle:2017tfw}   \\
      \hline
      \multirow{2}*{$\Lcp\to p\eta'$}         & $0.56^{+0.25}_{-0.21}$              & BESIII(2022)\cite{BESIII:2022izy} & $\Lcp\to (pK^+K^-\pi^0)_\text{NR}$       & $<0.06$                & Belle(2017)\cite{Belle:2017tfw}   \\
      \cline{4-6}
      ~                                       & $0.47\pm0.10$                       & Belle(2022)\cite{Belle:2021vyq}   & \multirow{2}*{$\Lcp\to pK^+\pi^-$}       & $0.16\pm0.02$          & Belle(2016)\cite{Belle:2015wxn}   \\
      \cline{1-3}
      $\Lcp\to p\rho$                         & $1.52\pm0.44$                       & LHCb(2024)\cite{LHCb:2024hju}     & ~                                        & $0.10\pm0.01$          & LHCb(2018)\cite{LHCb:2017xtf}     \\
      \hline
      \multirow{4}*{$\Lcp\to p\omega$}        & $0.94\pm0.39$                       & LHCb(2018)\cite{LHCb:2017yqf}     & ~                                        &                        &                                   \\
      ~                                       & $0.83\pm0.11$                       & Belle(2021)\cite{Belle:2021btl}   & ~                                        &                        &                                   \\
      ~                                       & $1.11\pm0.21$                       & BESIII(2023)\cite{BESIII:2023ooh} & ~                                        &                        &                                   \\
      ~                                       & $0.98\pm0.31$                       & LHCb(2024)\cite{LHCb:2024hju}     & ~                                        &                        &                                   \\
      \cline{1-3}
      $\Lcp\to p\phi$                         & $1.06\pm0.22$                       & BESIII(2016)\cite{BESIII:2016ozn} & ~                                        & ~                      & ~                                 \\
      \hline\hline
      \multicolumn{6}{l}{\textbf{$\Lambda$-involved}}                                                                                                                                                            \\ \hline
      \multirow{2}*{$\Lcp\to \Lambda K^+$}    & $0.62\pm0.06$                       & BESIII(2022)\cite{BESIII:2022tnm} & \multirow{2}*{$\Lcp\to\Lambda K^+\pi^0$} & $<2.0$                 & BESIII(2024)\cite{BESIII:2023dvx} \\
      ~                                       & $0.66\pm0.04$                       & Belle(2023)\cite{Belle:2022uod}   & ~                                        & $1.49\pm0.29$          & BESIII(2024)\cite{BESIII:2023sdr} \\
      \hline
      \multirow{3}*{$\Lcp\to \Lambda K^{*+}$} & $2.40\pm0.59$($\theta_0=0^\circ$)   & BESIII(2025)\cite{BESIII:2024xny} & $\Lcp\to \Lambda K^0_S \pi^+$            & $1.73\pm0.29$          & BESIII(2025)\cite{BESIII:2024xny} \\
      \cline{4-6}
      ~                                       & $5.21\pm0.75$($\theta_0=109^\circ$) & BESIII(2025)\cite{BESIII:2024xny} & $\Lcp\to \Lambda K^+\pi^+\pi^-$          & $0.41\pm0.15$          & BESIII(2024)\cite{BESIII:2023sdr} \\
      \cline{4-6}
      ~                                       & $1.29\pm0.44$($\theta_0=221^\circ$) & BESIII(2025)\cite{BESIII:2024xny} & ~                                        & ~                      & ~                                 \\
      \hline\hline
      \multicolumn{6}{l}{\textbf{$\Sigma$-involved}}                                                                                                                                                             \\ \hline
      \multirow{2}*{$\Lcp\to \Sigma^0K^+$}    & $0.47\pm0.10$                       & BESIII(2022)\cite{BESIII:2022wxj} & $\Lcp\to \Sigma^+K^+\pi^-$               & $2.00\pm0.28$          & BESIII(2023)\cite{BESIII:2023rky} \\
      \cline{4-6}
      ~                                       & $0.36\pm0.03$                       & Belle(2023)\cite{Belle:2022uod}   & $\Lcp\to \Sigma^+K^+\pi^-\pi^0$          & $<0.01$                & BESIII(2023)\cite{BESIII:2023rky} \\
      \hline
      $\Lcp\to \Sigma^+K^0_S$                 & $0.48\pm0.14$                       & BESIII(2022)\cite{BESIII:2022wxj} & \multirow{2}*{$\Lcp\to\Sigma^0K^+\pi^0$} & $<1.8$                 & BESIII(2024)\cite{BESIII:2023dvx} \\
      \cline{1-3}
      ~                                       & ~                                   & ~                                 & ~                                        & $<0.50$                & BESIII(2024)\cite{BESIII:2025vlo} \\
      \cline{4-6}
      ~                                       & ~                                   & ~                                 & $\Lcp\to \Sigma^0K^+\pi^+\pi^-$          & $<0.65$                & BESIII(2024)\cite{BESIII:2025vlo} \\
      \cline{4-6}
      ~                                       & ~                                   & ~                                 & $\Lcp\to \Sigma^-K^+\pi^+$               & $0.38\pm0.12$          & BESIII(2024)\cite{BESIII:2023iwu} \\
      \hline \hline
    \end{tabular}
  \end{center}
\end{table*}

\vspace{0.2cm}
\subsubsection{Decay asymmetry measurement}
\vspace{0.2cm}

Polarization parameters are a set of physical observables that characterize the parity violation in weak decay.
They were introduced in 1957 by T. D. Lee and C. N. Yang~\cite{Lee:1957qs}, and are therefore also referred to as the Lee-Yang parameters.
For instance, in a weak decay $\Lcp\to BP$ ($B$ denotes a $J^P =\frac{1}{2}^+$ baryon and $P$ denotes a $J^P =0^-$ pseudoscalar meson), polarization parameters can be defined as $\alpha_{BP}\equiv\text{Re}(\mathcal{S}^*\mathcal{P})/(|\mathcal{S}|^2+|\mathcal{P}|^2)$, $\beta_{BP}\equiv\text{Im}(\mathcal{S}^*\mathcal{P})/(|\mathcal{S}|^2+|\mathcal{P}|^2)$, and $\gamma_{BP}\equiv(|\mathcal{S}|^2-|\mathcal{P}|^2))/(|\mathcal{S}|^2+|\mathcal{P}|^2)$, where $\mathcal{S}$ and $\mathcal{P}$ stand for the parity-violated and parity-conserving waves, and they satisfy $\alpha^2_{BP}+\beta^2_{BP}+\gamma^2_{BP}=1$.
The effects of parity violation are mainly determined by studying the angular distributions of the produced daughter baryon $B$ in the rest frame of the charmed baryon.
Experimentally, the information of the polarization of the final-state baryon would enhance the sensitity of accessing the parity violation parameters, such as the angular analysis of the daughter baron weak decays.

The BESIII, Belle(II), and LHCb experiments possess distinct characteristics and technical advantages in measuring the polarization parameters.
For BESIII, the charmed baryon $\Lcp$ is produced in pairs with the anti-particle $\Lcm$ via virtual photon exchange in $\ee$ annihilations, which induces a unique feature of quantum interference between the $\Lcp$ and $\Lcm$ production amplitudes.
This interference leads to a non-zero production transverse polarization, which is a direct consequence of the interference between the two production amplitudes~\cite{BESIII:2019odb,BESIII:2025zbz}, whose sizes are energy-dependent.
While the production cross section reflects the magnitude of the electromagnetic form factors, the transverse polarization offers sensitivity to their relative phase, thereby providing complementary information on the dynamics of $\Lcp\Lcm$ pair production via virtual photon exchange.
The Belle and Belle~II experiments primarily study the charmed baryons through the continuum production process.
Owing to their large-statsitcs data samples, 
one-dimensional angular fits already present very good sensitivity to the polarization parameters.
For instance, in the decay $\Lcp\to\Lambda\pi^+$, the product of the polarization parameters $\alpha^{\Lcp}_{\Lambda\pi^+}\alpha^{\Lambda}_{p\pi^-}$ can be determined by fitting to the distribution of the helicity angle between the proton direction and the opposite of the $\Lcp$ momentum in the $\Lambda$ rest frame. In the context, the superscripts in the polarization parameters denote the parent baryon.
Given an external input of the previously measured value of $\alpha^{\Lambda}_{ p\pi^-}$, the value of $\alpha^{\Lcp}_{\Lambda\pi^+}$ can then be calculated.
The LHCb experiment, primarily known as a $b$-quark factory, measures the $\Lcp$ polarization parameters by studying the $\Lcp$ produced from the $\Lambda^0_b$ weak decays, such as $\Lambda^0_b\to\Lcp\pi^-$ and $\Lambda^0_b\to\Lambda^+_c\mu^-X$ , in which the produced $\Lcp$ baryons have sizable longitudinal polarizations through the $P$-violating decays of the parent $\Lambda^0_b$~\cite{LHCb:2024tnq}.
This enables LHCb to perform high-precision measurements of polarization observables.
Nonetheless, relatively poor photon detection at LHCb restricts its sensitivity in the decay modes involving photons.
Such processes are therefore best studied at BESIII or Belle (II), where photon showers in the electromagnetic calorimeter are less contaminated due to the accompanying hadron showers.
Table~\ref{tab:LcHad_PO} summarizes the polarization parameters of the $\Lcp$ decays measured in the BESIII, Belle and LHCb experiments.

\begin{table*}[tbp]
  \caption{The determined polarization parameters of various $\Lcp$ decay modes.} \label{tab:LcHad_PO}
  \begin{center}
    \footnotesize
    \begin{tabular}{lcc|lcc}
      \hline\hline
      Mode    &  $\alpha$        & Experiment   & Mode       &  $\alpha$ & Experiment                        \\
      \hline
      \multicolumn{3}{l|}{\textbf{Nucleon-involved}}                  & $\Lcp\to \Lambda(1600)\pi^+$ & $0.2\pm0.5$                       & LHCb(2023)\cite{LHCb:2022sck}                                                                                             \\
      \hline
      \multirow{2}*{$\Lcp\to pK_S^0$}                                & $-0.75\pm0.10$               & LHCb(2024)\cite{LHCb:2024tnq}    & \multirow{2}*{$\Lcp\to \Lambda(1670)\pi^+$}                   & $0.82\pm0.08$         & LHCb(2023)\cite{LHCb:2022sck}     \\
      ~                & $-0.92^{+0.14}_{-0.09}$                & BESIII(2025)\cite{BESIII:2025zbz}                                                  & ~                                                             & $0.21\pm0.43$         & BESIII(2025)\cite{BESIII:2024mbf} \\
      \hline
      $\Lcp\to p\bar{K}^*_0(700)^0$                                   & $-0.1\pm0.7$                 & LHCb(2023)\cite{LHCb:2022sck}      & $\Lcp\to \Lambda(1690)\pi^+$                                  & $0.958\pm0.034$       & LHCb(2023)\cite{LHCb:2022sck}     \\
      \hline
      $\Lcp\to p\bar{K}^*_0(892)^0$                                   & $0.87\pm0.03$                 & LHCb(2023)\cite{LHCb:2022sck}      & $\Lcp\to \Lambda(2000)\pi^+$                                  & $-0.57\pm0.19$        & LHCb(2023)\cite{LHCb:2022sck}     \\
      \hline
 
      $\Lcp\to p\bar{K}^*_0(1430)^0$                                  & $0.34\pm0.14$                & LHCb(2023)\cite{LHCb:2022sck}     & \multicolumn{3}{l}{\textbf{$\Sigma$-involved}}                                                             \\
      \hline
     $\Lcp\to \Delta(1232)^{++}K^-$                                  & $0.55\pm0.04$                & LHCb(2023)\cite{LHCb:2022sck}       & \multirow{2}*{$\Lcp\to \Sigma^+\pi^0$}                           & $-0.48\pm0.03$        & Belle(2023)\cite{Belle:2022bsi}                 \\
      \cline{1-3}
            $\Lcp\to \Delta(1600)^{++}K^-$                                  & $-0.50\pm0.18$               & LHCb(2023)\cite{LHCb:2022sck}   & ~                                           & $-0.59\pm0.05$        & BESIII(2025)\cite{BESIII:2025zbz}  \\
      \hline
      $\Lcp\to \Delta(1700)^{++}K^-$                                  & $0.22\pm0.08$                & LHCb(2023)\cite{LHCb:2022sck}   & $\Lcp\to \Sigma^+\eta$       & $-0.99\pm0.06$                    & Belle(2023)\cite{Belle:2022bsi}                                                                                           \\
      \hline
       \multicolumn{3}{l|}{\textbf{$\Lambda$-involved}}  & $\Lcp\to \Sigma^+\eta'$                                       & $-0.46\pm0.07$        & Belle(2023)\cite{Belle:2022bsi}   \\ \hline
      \multirow{3}*{$\Lcp\to \Lambda\pi^+$}                                                                       & $-0.785\pm0.007$             & LHCb(2024)\cite{LHCb:2024tnq}   
       & \multirow{2}*{$\Lcp\to \Sigma^0\pi^+$}                                   & $-0.46\pm0.02$        & Belle(2023)\cite{Belle:2022uod}        \\
            ~                                                               & $-0.755\pm0.006$             & Belle(2023)\cite{Belle:2022uod}    & ~                                           & $-0.50\pm0.08$        & BESIII(2025)\cite{BESIII:2025zbz}   \\
      \cline{4-6}

      ~       & $-0.790\pm0.033$               & BESIII(2025)\cite{BESIII:2025zbz} 
       & $\Lcp\to \Sigma(1385)^+\pi^0$                                 & $-0.917\pm0.089$      & BESIII(2022)\cite{BESIII:2022udq} \\
      \hline
      
      \multirow{2}*{$\Lcp\to \Lambda K^+$}                            & $-0.59\pm0.05$               & Belle(2023)\cite{Belle:2022uod}  & $\Lcp\to \Sigma(1385)^+\eta$                                  & $-0.61\pm0.16$        & BESIII(2025)\cite{BESIII:2024mbf} \\
      \cline{4-6}
      ~                                                               & $-0.52\pm0.05$               & LHCb(2024)\cite{LHCb:2024tnq}    & $\Lcp\to \Sigma(1385)^0\pi^+$                                 & $-0.79\pm0.11$      & BESIII(2022)\cite{BESIII:2022udq} \\
      \hline

      $\Lcp\to \Lambda\rho(770)^+$                                    & $-0.763\pm0.070$             & BESIII(2022)\cite{BESIII:2022udq}    & $\Lcp\to \Sigma^0 K^+$                                        & $-0.54\pm0.20$        & Belle(2023)\cite{Belle:2022uod}   \\
      \hline
      $\Lcp\to \Lambda a(980)^+$                                      & $-0.91^{+0,20}_{-0.12}$      & BESIII(2025)\cite{BESIII:2024mbf}   & \multicolumn{3}{l}{\textbf{$\Xi$-involved}}                                                                \\
      \hline
      $\Lcp\to \Lambda(1405)\pi^+$                                    & $0.58\pm0.28$                & LHCb(2023)\cite{LHCb:2022sck}       & $\Lcp\to \Xi^0K^+$                                            & $0.01\pm0.16$         & BESIII(2024)\cite{BESIII:2023wrw} \\
      \hline 
            
      $\Lcp\to \Lambda(1520)\pi^+$                                    & $0.93\pm0.09$                & LHCb(2023)\cite{LHCb:2022sck} & & & \\ \hline \hline
     Mode    &  $\beta$        & Experiment   & Mode       &  $\gamma$ & Experiment                        \\
      \hline
\multirow{2}*{$\Lcp\to \Lambda\pi^+$}   & $0.378\pm0.015$               & LHCb(2024)\cite{LHCb:2024tnq}    & \multirow{2}*{$\Lcp\to \Lambda\pi^+$}   & $0.491\pm0.012$        & LHCb(2024)\cite{LHCb:2024tnq}      \\
~    & $0.37^{+0.17}_{-0.25}$       & BESIII(2025)\cite{BESIII:2025zbz}   & ~  & $0.64^{+0.10}_{-0.20}$        & BESIII(2025)\cite{BESIII:2025zbz} \\ \hline
$\Lcp\to \Sigma^0 \pi^+$     & $0.70^{+0.14}_{-0.48}$               & BESIII(2025)\cite{BESIII:2025zbz}   & $\Lcp\to \Sigma^0 \pi^+$         & $-0.50^{+0.59}_{-0.30}$        & BESIII(2025)\cite{BESIII:2025zbz} \\ \hline
$\Lcp\to \Sigma^+ \pi^0$     & $0.76^{+0.05}_{-0.24}$               & BESIII(2025)\cite{BESIII:2025zbz}   & $\Lcp\to \Sigma^+ \pi^0$         & $-0.26^{+0.48}_{-0.38}$          & BESIII(2025)\cite{BESIII:2025zbz} \\ \hline
$\Lcp\to \Xi^0 K^+$     & $-0.64\pm0.70$               & BESIII(2024)\cite{BESIII:2023wrw}   & $\Lcp\to \Xi^0 K^+$         & $-0.77\pm0.59$        & BESIII(2024)\cite{BESIII:2023wrw}  \\ \hline
$\Lcp\to \Lambda K^+$     & $0.33\pm0.08$               & LHCb(2024)\cite{LHCb:2024tnq}   & $\Lcp\to \Lambda K^+$         & $-0.799\pm0.041$        & LHCb(2024)\cite{LHCb:2024tnq}  \\ \hline \hline
    \end{tabular}
  \end{center}
\end{table*}

Based on about 100,000 $\Lambda_c^+\bar{\Lambda}_c^-$ pairs produced near threshold at $4.6$~GeV at BESIII~\cite{BESIII:2017kqg}, a joint angular  analysis has been implemented 
simultanously for the four decay modes $\Lambda_c^+ \to pK_S^0$, $\Lambda\pi^+$, $\Sigma^0\pi^+$ and $\Sigma^+\pi^0$~\cite{BESIII:2019odb} and the polarization parameters $\alpha$, $\beta$ and $\gamma$ were measured, except those of $\beta^{\Lcp}_{pK^0_S}$ and $\gamma^{\Lcp}_{pK^0_S}$, as no information on the polarization of the proton is available. Especially, the parameters $\alpha^{\Lcp}_{pK^0_S}$ and $\alpha^{\Lcp}_{\Sigma^0\pi^+}$ are determined for the first time to be $0.18\pm0.43\pm0.14$ and $-0.73\pm0.17\pm0.07$, respectively.
The result shows that $\alpha^{\Lcp}_{\Sigma^+\pi^0}$ differs from the positive predictions by at least $8\sigma$, and rules out those model calculations~\cite{Cheng:1993gf,Xu:1992vc,Korner:1992wi,Cheng:1991sn,Ivanov:1997ra,Zenczykowski:1993jm}.
In addition, the study shows that no model gives predictions fully consistent with all the results of the four polarization parameters.
Later on, Belle improved precisions on $\alpha^{\Lcp}_{\Lambda \pi^+}$, $\alpha^{\Lcp}_{\Sigma^0\pi^+}$ and $\alpha^{\Lcp}_{\Sigma^+\pi^0}$, with one-dimensional angular analysis in the incluive $\Lcp$ production~\cite{Belle:2022uod,Belle:2022bsi} based on a data sample of $980\,\ipb$ collected at or near the $\Upsilon(nS)$ ($n$=1, 2, 3, 4 and 5) resonances. Meanwhile, the polarization parameters $\alpha^{\Lcp}_{\Sigma^+ \eta^{(\prime)}}$ $\alpha^{\Lcp}_{\Lambda K^+}$ and $\alpha^{\Lcp}_{\Sigma^0 K^+}$ are measured for the first time at Belle~\cite{Belle:2022uod,Belle:2022bsi}.
In this study, as only one-dimensional angular fit was performed, the parameters of $\beta$ and $\gamma$ are not accessible, since many sub-level angular variables are integrated out.
In 2023, LHCb performed a simultanous multi-dimensional angular analysis of the cascade decays $\Lambda^0_b \to \Lambda_c^+ h^-$ ($h$=$\pi$, $K$) followed with $\Lcp \to \Lambda h^+$ with $\Lambda \to p \pi^-$ or $\Lambda_c^+ \to pK_S^0$ with high statistics samples based on a data sample of proton-proton collisions of 9\,\ifb at center-of-mass energies of 7, 8, and 13 TeV~\cite{LHCb:2024tnq}.
Similar to the BESIII multi-dimensional fit, besides the polarization parameter $\alpha$, the parameters $\beta$ and $\gamma$ for $\Lambda_c^+ \to \Lambda\pi^+$ and $\Lambda K^+$ can be probed. In this LHCb analysis, the result of $\alpha^{\Lcp}_{\Lambda\pi^+}$ is $-0.785\pm0.006\pm0.003$, which deviates from the Belle result of $-0.755\pm0.005\pm0.003$ by about $4\sigma$. In 2025, BESIII updated the measurements of the polarization parameters for $\Lcp \to pK_S^0$, $\Lambda\pi^+$, $\Sigma^0\pi^+$ and $\Sigma^+\pi^0$ with a larger data sample of 6.4\,\ifb collected at center-of-mass energies between 4.60 and 4.95\,GeV~\cite{BESIII:2025zbz}, which suppresses the previous measurement in 2019~\cite{BESIII:2019odb} with much improved precisions, as listed in Table~\ref{tab:LcHad_PO}. The obtained  $\alpha^{\Lcp}_{\Lambda\pi^+}$ is $-0.790\pm0.032\pm0.009$, which is closer to the LHCb result within $1\sigma$.
In addition, non-zero transverse polarization of the $\Lcp$ produced in $\ee\to\Lcp\Lcm$ is observed for the first time~\cite{BESIII:2025zbz}.

For multibody decays, PWA is a crucial method to obtain polarization information of the involved two-body resonant processes, such as the analyses of $\Lcp\to\Lambda\pi^+\pi^0$~\cite{BESIII:2022udq} and $\Lcp\to\Lambda\pi^+\eta$~\cite{BESIII:2024mbf} carried out at BESIII.
Polarization information of the involved intermediate processes can be accessed according to the multi-dimensional angular distributions in the final states.
Hence, polarization parameters of the $\Lcp$ decays into excited states are determined for the first time, such as $\alpha^{\Lcp}_{\Lambda\rho(770)^+}$, $\alpha^{\Lcp}_{\Lambda a(980)^+}$, $\alpha^{\Lcp}_{\Sigma(1385)\pi}$ and $\alpha^{\Lcp}_{\Sigma(1385)^+\eta}$.
In a similar fashion, LHCb performed PWA of $\Lcp\to pK^-\pi^+$ in 2023~\cite{LHCb:2022sck} and extract polarization information for the intermediate processes, such as $\Lcp\to p \bar{K}^{*}$, $\Delta^{*} K^-$ and $\Lambda^{*}\pi^+$. 
The numerical values of the polarization parameters can be found in Table~\ref{tab:LcHad_PO}.
The theoretical study of polarization parameters for final states involving excited states is quite challenging.
Therefore, the experimental measurements provide ciritical information for improving theoretical models.

In the last century, various theoretical predictions showed the polarization parameter $\alpha_{\Xi^0 K^+}$ should be zero~\cite{Korner:1992wi,Xu:1992vc,Zenczykowski:1993jm,Ivanov:1997ra,Sharma:1998rd} since the vanished $s$-wave amplitude. However, in recent years, many theoretical models give non-zero predictions and  some even reach the physical postive boundary~\cite{Geng:2019xbo,Zou:2019kzq,Zhong:2022exp}.
Under such circumstances, the measurement of $\alpha^{\Lcp}_{\Xi^0 K^+}$ proves to be especially crucial and indispensable.
In 2024, a 7-dimensional angular distribution analysis of the three-level weak decay process $\Lcp\to\Xi^0 K^+$ was conducted with 378 events by BESIII, and led to the successful extraction of a set of polarization parameters~\cite{BESIII:2023wrw}.
The helicity angle definition and angular fitting results are shown in Figure~\ref{fig:LcXi0K}, which allowed for the extraction of polarization parameters in this pure $W$-exchange process.
The polarization parameters are determined to be $\alpha^{\Lcp}_{\Xi^0 K^+}=0.01\pm 0.16\pm0.03$,
$\beta^{\Lcp}_{\Xi^0 K^+}=-0.64\pm 0.69\pm 0.13$ and
$\gamma^{\Lcp}_{\Xi^0 K^+}=-0.77\pm 0.58\pm 0.11$.
Considering the known BF and the determined $\beta^{\Lcp}_{\Xi^0 K^+}$, decay dynamic parameters, $|A|$ and $|B|$ are derived into two sets of solutions: the first one is $|A|_1=1.6^{+1.9}_{-1.6}\pm0.4$ and $|B|_1=18.3\pm2.8\pm0.7$, and the second one is $|A|_2=4.3\pm0.7\pm0.2$ and $|B|_2=6.7^{+8.3}_{-6.7}\pm1.6$.
Moreover, the strong  phase shift was extracted as $\delta_p-\delta_s=-1.55\pm0.25\pm0.05$ rad or $1.59\pm0.25\pm0.05$ rad~\cite{Wang:2024wrm}, which exerts a dominant influence on the polarization parameter $\alpha^{\Lcp}_{\Xi^0 K^+}$.
After taking into account the correlation between $\alpha^{\Lcp}_{\Xi^0 K^+}$ and $\beta^{\Lcp}_{\Xi^0 K^+}$ as discussed in Ref.~\cite{Wang:2024wrm}, the data tend to favor the solution with the negative strong phase shift, i.e., $\delta_p-\delta_s=-1.55\pm0.25\pm 0.05$ rad.
These new observables present a novel direction in testing theoretical calculations~\cite{Duan:2024zjv,Wang:2024qff,Cheng:2024lsn,Jia:2024pyb,Shi:2024plf,Sun:2024mmk,He:2024pxh,Zhong:2024zme,Cheng:2024rdi,Geng:2023pkr}. In particlur, this study reveals that the realtive phases between diffenent amplitudes is vital, which was not sensitive in the BF test and hence, not previously fully validated  in most theoretical models.

\begin{figure}[tbp]\centering
  \includegraphics[width=0.47\linewidth]{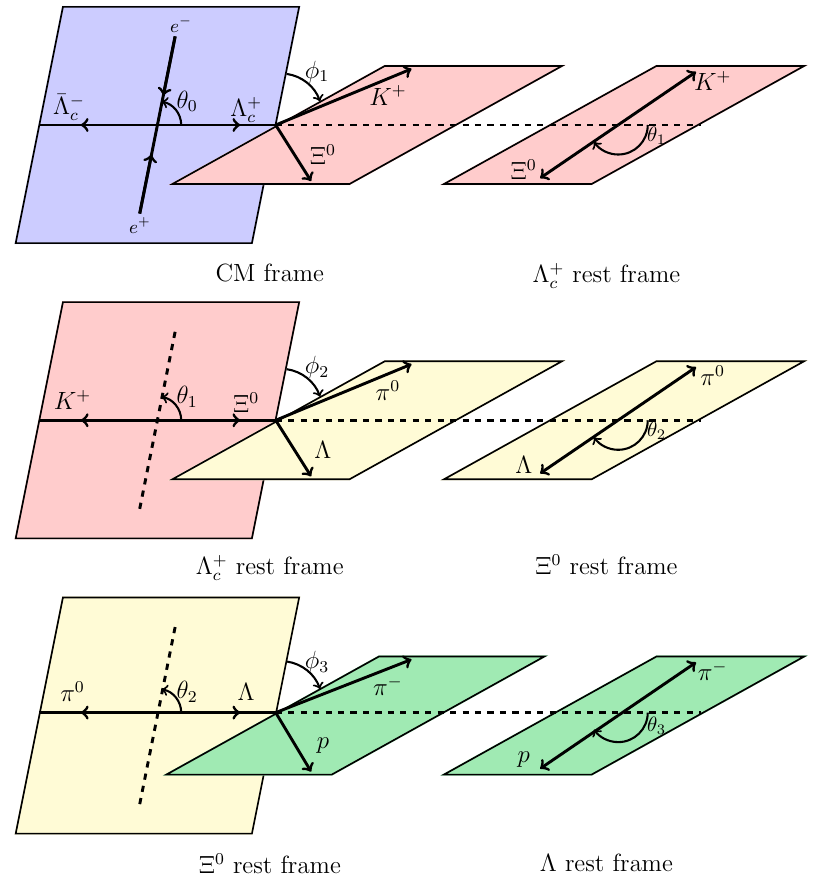}
  \includegraphics[width=0.4\linewidth]{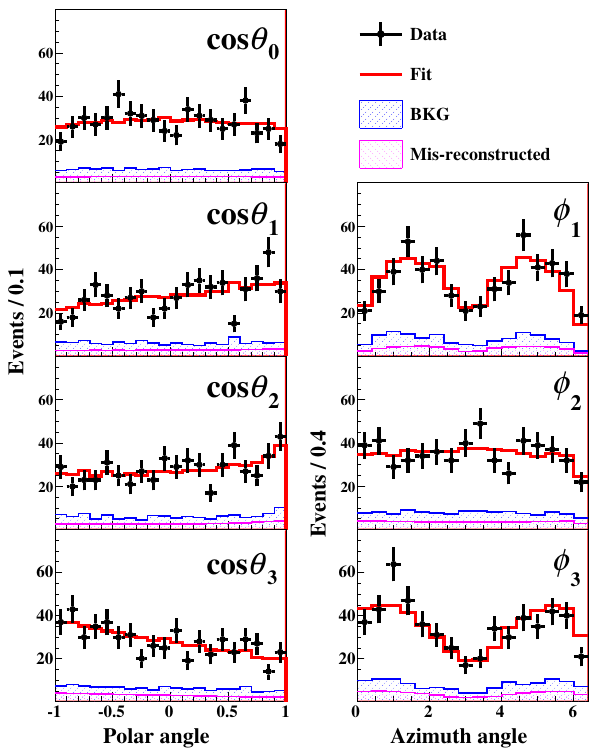}
  \caption{(left) Definitions of the helicity frames and related angles for $e^{+}e^{-}\to\Lambda_{c}^{+}\bar{\Lambda}_c^{-}, \Lambda_{c}^{+}\to\Xi^{0}K^{+}, \Xi^{0}\to\Lambda\pi^0$, and $\Lambda\to p\pi^{-}$. (right) Projections of the best fit onto different angular distributions. Black points with error bars are data, red solid lines represent the fitting results.}
  \label{fig:LcXi0K}
\end{figure}

The $\Lcp$ CS decays has relatively low yields and present a significant experimental challenge for polarization parameter measurements.
As aforementioned studies at Belle in Refs.~\cite{Belle:2022uod,Belle:2022bsi} the polarization parameters of the CS decays $\alpha^{\Lcp}_{\Lambda K^+}=-0.585\pm0.049\pm0.018$ and $\alpha^{\Lcp}_{\Sigma^0K^+}=-0.54\pm0.18\pm0.09$~\cite{Belle:2022uod} for $\Lcp\to\Lambda K^+$ and $\Lcp\to\Sigma^0 K^+$ are determined, for the first time. While in the mutli-dimensional angular analysis at LHCb in Ref.~\cite{LHCb:2024tnq}, the determined $\alpha^{\Lcp}_{\Lambda K^+}=-0.516\pm 0.041\pm0.021$ is in agreement with that previously obtained by Belle.
In addition, LHCb also simultanously measured the polarization parameters of $\beta^{\Lcp}_{\Lambda K^+}$ and $\gamma^{\Lcp}_{\Lambda K^+}$ for the first time.
As can be seen in Table~\ref{tab:LcHad_PO}, polarization measurements for the CS decays remain limited. Nevertheless, such CS decays are of particular interest in searching for $C\!P$ violation, as in the SM  the interference bewteen the tree and loop diagrams 
automatically leads to an effect of $C\!P$
asymmetry.
Experimental test on the inequalities of $\bar{\alpha}_{\bar{B}\bar{P}}\neq-\alpha_{BP}$ and $\bar{\beta}_{\bar{B}\bar{P}}\neq-\beta_{BP}$ would indicate $C\!P$ asymmetry, where the overline denotes the anti-particle.
The asymmetry observables $A^{\alpha}_{C\!P}(BP)\equiv(\alpha_{BP}+\bar{\alpha}_{\bar{B}\bar{P}})/(\alpha_{BP}-\bar{\alpha}_{\bar{B}\bar{P}})$ and $A^{\beta}_{C\!P}(BP)\equiv(\beta_{BP}+\bar{\beta}_{\bar{B}\bar{P}})/(\alpha_{BP}-\bar{\alpha}_{\bar{B}\bar{P}})$ 
can be defined, which provide a novel way in testing $C\!P$ symmetry besides the observable of decay rate asymmetry.
Such $C\!P$ asymmetries in CS decay modes were firstly tested via $\Lcp\to\Lambda K^+$ and $\Lcp\to\Sigma^0 K^+$ at Belle~\cite{Belle:2022uod}, which gives
$A^{\alpha}_{C\!P}(\Lambda K^+)=-0.023\pm0.086\pm0.071$ and $A^{\alpha}_{C\!P}(\Sigma^0 K^+)=0.08\pm0.35\pm0.14$. 
LHCb searched for the $C\!P$ asymmetry in $\Lcp\to\Lambda K^+$~\cite{LHCb:2024tnq} and found $A^{\alpha}_{C\!P}(\Lambda K^+)=0.102\pm0.080\pm0.023$ and 
$A^{\beta}_{C\!P}(\Lambda K^+)=-0.04\pm0.15\pm0.02$.
With the current experimental precisions, no evidence for polarization-induced $C\!P$ asymmetry in charmed baryon decays are seen . Future measurements with substantially larger data samples and improved detector performance will be essential to improve the $C\!P$ test sensitivities.
 
\subsection{$\Xi^0_c$ and $\Xi^+_c$ decays}

In recent charmed baryon studies, there were several major breakthroughs in $\Xi_{c}^{0,+}$ decays.
Before 2019, no absolute BFs for their decays have been measured\cite{ParticleDataGroup:2018ovx}. Only the BF ratios realtive to $\BR(\Xi_{c}^0\to \Xi^-\pi^+)$ and $\BR(\Xi_{c}^+ \to \Xi^-\pi^+ \pi^+)$ are determined directly in experiments.
In 2019, Belle carried out the first BFs for the  $\Xi_{c}^{0,+}$ decays.
Based on the data sample with $(772\pm11)\times10^6$ $B\bar{B}$ pairs collected at $\Upsilon(4S)$ resonance, Belle reported the first measurement of the absolute BF $\mathcal{B}(\Xi_{c}^0\to\Xi^-\pi^+)=(1.80\pm0.50\pm0.14)\%$~\cite{Belle:2018kzz}, where the $\Xcz$  inclusive decays are tagged via the process
$e^+e^-\to B^+ B^-$, $B^-\to \Lcm\Xcz$. In the same studies, the absolute BFs for the other two CF decays are also obtained to be $\mathcal{B}(\Xi_c^0\to pK^-K^-\pi^+)=(0.58\pm0.23\pm0.05)\%$ and $\mathcal{B}(\Xi_c^0\to\Lambda K^-\pi^+)=(1.17\pm0.37\pm0.09)\%$  at Belle~\cite{Belle:2018kzz}.
In a similar fashion, Belle implmented the measurement of the absolute BFs for $\Xi_{c}^+$ decays for the first time, which gives $\mathcal{B}(\Xi_{c}^{+}\to\Xi^-\pi^+\pi^+)=(2.86\pm1.21\pm0.38)\%$ and $\mathcal{B}(\Xi_{c}^+\to pK^-\pi^+)=(0.45\pm0.21\pm0.07)\%$ \cite{Belle:2019bgi}, by tagging the $\Xi_{c}^+$ inclusive decays via the process $e^+e^-\to B^0 \bar{B}^0$, $\bar{B}^0\to \Lcm\Xi_{c}^+$.
Therefore, the BFs for other $\Xi_{c}$ decays can be derived from their determined BF ratios relative to $\mathcal{B}(\Xi_{c}^0\to\Xi^-\pi^+)$ or $\mathcal{B}(\Xi_{c}^+\to\Xi^-\pi^+\pi^+)$.
So far, the precisions of the absolute BFs of $\Xi_c$ decays are still low, compared to the BF of the decay $\Lambda_{c}^+\to pK^-\pi^+$~\cite{Belle:2013jfq}.
Later, LHCb measured the BF for $\Xi_{c}^+\to pK^-\pi^+$ as $\mathcal{B}(\Xi_{c}^+\to pK^-\pi^+)=(1.135\pm0.002\pm0.387)\%$ \cite{LHCb:2020gge}, which is dominated by systematic uncertainties of 
production fractions of $f_{\Xi_{c}^0}/f_{\Xi_{c}^+}$ and $f_{\Xi_{c}^0}/f_{\Lambda_{c}^+}$.
Overall, $\mathcal{B}(\Xi_{c}^+\to pK^-\pi^+)$ measured by LHCb is moderately higher, but in agreement with the value reported in the Belle measurement. Future measurements with improved precisions are needed to clarify the potential discrepancy.

\begin{table*}[tbp]
  \caption{The recently measured BFs for the $\Xi^+_c$ and $\Xi^0_c$ decays~(in units of \%). Values marked with $\dagger$ are mutliplied with the reference mode in PDG~\cite{ParticleDataGroup:2024cfk}.} \label{tab:XicHad_BF}
  \begin{center}
    \begin{tabular}{lcc|lcc}
      \hline\hline
      $\Xcz$ mode                                  & BF    & Experiment                          & $\Xcp$ mode                                          &  BF     & Experiment                             \\
      \hline
      $\Lambda K^0_S$            & $0.33\pm0.08$ & Belle(2022)\cite{Belle:2021avh} 
        &      $\Xi^-\pi^+\pi^+$                  & $2.86\pm1.27$   & Belle(2019)\cite{Belle:2019bgi}     \\

      $\Lambda\bar{K}^{*0}$      & $0.33\pm0.11$ & Belle(2021)\cite{Belle:2021zsy}  &
 $\Sigma^+K^0_S$                    & $0.19\pm0.09$   & Belle(II)(2025)\cite{Belle-II:2025klu} \\   \cline{1-3}      

      $\Sigma^+ K^-$             & $0.18\pm0.04$ & Belle(2022)\cite{Belle:2021avh}   &
       $\Xi^0\pi^+$                       & $0.72\pm0.32$   & Belle(II)(2025)\cite{Belle-II:2025klu} \\ \cline{4-6}

      $\Sigma^+K^{*-}$           & $0.61\pm0.21$ & Belle(2021)\cite{Belle:2021zsy}   
            & $pK^0_S$                        & $0.07\pm0.03$   & Belle(II)(2025)\cite{Belle:2024xcs}    \\

       $\Sigma^0\bar{K}^{*0}$     & $1.24\pm0.37$ & Belle(2021)\cite{Belle:2021zsy}   &  $\Lambda\pi^+$                    & $0.05\pm0.02$   & Belle(II)(2025)\cite{Belle:2024xcs}     \\

   $\Sigma^0 K^0_S$           & $0.05\pm0.02$ & Belle(2022)\cite{Belle:2021avh}   &  $\Sigma^0\pi^+$      & $0.12\pm0.06$   & Belle(II)(2025)\cite{Belle:2024xcs}    \\

     $\Xi^0\pi^0$               & $0.69\pm0.14$ & Belle(II)(2024)\cite{Belle:2024ikp}  &  $\Xi^0K^+$                      & $0.05\pm0.02$   & Belle(II)(2025)\cite{Belle-II:2025klu}  \\ \cline{4-6}

       $\Xi^0\eta$                & $0.16\pm0.04$ & Belle(II)(2024)\cite{Belle:2024ikp}  & 
\multirow{2}*{$pK^-\pi^+$}         &   
      $0.45\pm0.22$   & Belle(2019)\cite{Belle:2019bgi}           \\
       
$\Xi^0\eta^{\prime}$               & $0.12\pm0.04$ & Belle(II)(2024)\cite{Belle:2024ikp} & ~                                             & 
      $1.14\pm0.39$   & LHCb(2020)\cite{LHCb:2020gge}    \\
      \cline{4-6}

 $\Xi^-\pi^+$               & $1.80\pm0.52$ & Belle(2019)\cite{Belle:2018kzz}    &  
 $p\phi$                        & $0.012\pm0.006^\dagger$ & LHCb(2019)\cite{LHCb:2019nxp}         \\ \cline{4-6}

 $pK^-K^-\pi^+$                     & $0.58\pm0.24$   & Belle(2019)\cite{Belle:2018kzz}    &  ~    & ~   & ~      \\

  $\Lambda K^-\pi^+$                 & $1.17\pm0.38$   & Belle(2019)\cite{Belle:2018kzz}  & ~    & ~   & ~      \\        \cline{1-3}

      $\Xi^-K^+$                & $0.04\pm0.01^\dagger$ & Belle(2013)\cite{Belle:2013ntc}   & ~    & ~   & ~      \\

      $\Lambda \phi$  (CS)                 & $0.05\pm0.01^\dagger$   & Belle(2013)\cite{Belle:2013ntc}   & ~    & ~   & ~      \\

      $\Lambda (K^+K^-)_{\text{non-}\phi}$ & $0.04\pm0.01^\dagger$   & Belle(2013)\cite{Belle:2013ntc}     & ~    & ~   & ~      \\

       $\Xi^0 \phi $                      & $0.05\pm0.01^\dagger$   & Belle(2021)\cite{Belle:2020ito} & ~    & ~   & ~      \\ 
      
      $\Xi^0 (K^+K^-)_{\text{non-}\phi}$                      & $0.06\pm0.01^\dagger$   & Belle(2021)\cite{Belle:2020ito}     & ~    & ~   & ~      \\     \cline{1-3}
         
         \multirow{2}*{$\Lcp\pi^-$} & $0.55\pm0.18$ & LHCb(2020)\cite{LHCb:2020gge}    & ~    & ~   & ~      \\
         
    ~  & $0.54\pm0.14$ & Belle(2023)\cite{Belle:2022kqi}  & ~    & ~   & ~      \\ \hline \hline 
         

 \multicolumn{6}{c}{Intermediate resonances obtained in PWA of $\Xcp\to pK^-\pi^+$ at LHCb~\cite{LHCb:2025hul} }   \\ \hline
  Resonance & Fit Fraction (\%) & BF$^\dagger$  ($\times 10^{-4}$) & Resonance & Fit Fraction (\%) & BF$^\dagger$  ($\times 10^{-4}$) \\ \hline
$\Lz(1405)$ & $3.3\pm$1.5 &   $2.05\pm0.94$ &
$\Lz(1520)$ & 2.64$\pm$0.14  & $1.637\pm0.087$  \\
$\Lz(1600)$ & 2.0$\pm$1.7 & $  1.2\pm1.1$ &
$\Lz(1670)$ & 3.03$\pm$0.21 &  $1.88\pm0.14$ \\
$\Lz(1690)$ & 1.55$\pm$0.59 &  $0.96\pm0.37$ &
$\Lz(1710)$ & 2.3$\pm$1.9   &  $1.4\pm1.2$ \\
$\Lz(1800)$ & 1.48$\pm$0.61 & $0.92\pm0.38$ &
$\Lz(1810)$ & 1.3$\pm$1.0   & $0.83\pm0.63$  \\
$\Lz(1820)$ & 0.82$\pm$0.18 & $0.51\pm0.11$ &
$\Lz(1830)$ & 0.20$\pm$0.12   & $0.124\pm0.075$  \\  
$\Lz(1890)$ & 0.19$\pm$0.18 &  $0.12\pm0.11$ &
$\Lz(2000)$ & 7.4$\pm$1.4   &  $4.59\pm0.87$  \\
$\Kstarbzz(700)^0$ & 7.4$\pm$4.9 & $4.6\pm3.0$ &
$\Kstarb(892)^0$ & 28.6$\pm$1.2   &  $17.74\pm0.73$ \\
$\Kstarbzz(1430)^0$ & 15.6$\pm$7.4 &  $9.7\pm4.6$ &
$\Kstarbtwo(1430)^0$ & 3.3$\pm$2.8   &  $2.1\pm1.7$ \\
$\Delta(1232)^{++}$ & 17.2$\pm$1.4 &  $10.66\pm0.87$ &
$\Delta(1600)^{++}$ & 4.3$\pm$1.3   & $2.67\pm0.82$  \\
$\Delta(1620)^{++}$ & 3.3$\pm$1.0 &   $2.04\pm0.65$ &
$\Delta(1700)^{++}$ & 2.01$\pm$0.49   &  $1.25\pm0.31$ \\ \hline \hline
\end{tabular}

\end{center}
\end{table*}

Besides the absolute measurement of the BFs of the $\Xi_c$ decays, Belle (II) has carried out several relative measurements on the BFs of their CF and CS decays, where the BFs for the normalization modes are taken as input values. The relavent results are summarized in Table~\ref{tab:XicHad_BF}.
The BFs for the decays $\Xi_{c}^{0}\to\Lambda\bar{K}^{*0}$ and $\Lambda K^0_{S}$ are measured by taking the decay $\Xi_{c}^{0}\to\Xi^{-}\pi^+$ as reference channel, which give $\mathcal{B}(\Xi_{c}^{0}\to\Lambda\bar{K}^{*0})=(3.3\pm0.3\pm0.2\pm1.0)\times10^{-3}$ and $\mathcal{B}(\Xi_{c}^{0}\to\Lambda K^0_{S})=(3.27\pm0.11\pm0.17\pm0.73)\times10^{-3}$~\cite{Belle:2021zsy,Belle:2021avh}.
Belle also observed the decay modes with one $\Sigma^{0}$ or $\Sigma^{+}$ in the final states, including $\Xi_{c}^{0}\to\Sigma^{0}\bar{K}^{*0}$, $\Sigma^{+}K^{*-}$~\cite{Belle:2021zsy}, $\Sigma^{0}K^0_{S}$ and $\Sigma^+K^-$~\cite{Belle:2021avh}.
The BFs are calculated as $\mathcal{B}(\Xi_{c}^{0}\to\Sigma^{0}\bar{K}^{*0})=(12.4\pm0.05\pm0.05\pm0.36)\times10^{-3}$, $\mathcal{B}(\Xi_{c}^{0}\to\Sigma^{+}\bar{K}^{*-})=(6.1\pm1.0\pm0.4\pm1.8)\times10^{-3}$, $\mathcal{B}(\Xi_{c}^{0}\to\Sigma^{0}K^0_{S})=(0.54\pm0.09\pm0.06)\times10^{-3}$ and $\mathcal{B}(\Xi_{c}^{0}\to\Sigma^{+}K^{-})=(1.76\pm0.10\pm0.14\pm0.39)\times10^{-3}$.
We can see that the BF for $\Xi_{c}^{0}\to\Sigma^{0}\bar{K}^{*0}$ is much larger than that for $\Xi_{c}^0\to\Lambda\bar{K}^{*0}$, which conflicts with the SU(3) flavor symmetry and dynamical model predictions~\cite{Geng:2020zgr,Hsiao:2019yur,Korner:1992wi,Zenczykowski:1993jm}.

The relative BFs for the $W$-exchange-only decay $\Xi_{c}^{0}\to\Xi^{0} K^+K^-$ with the resonant polarized $\phi$ and the nonresonant decays were observed by Belle~\cite{Belle:2020ito}, with a total BF of about 0.11\%. Based on an analysis with azimuthally symmetric amplitude model, it was found that  $(48.1\pm4.2)\%$ decay resonantly through $\phi\to K^+K^-$, while $(51.9\pm4.2)\%$ decay directly to $\Xi^{0} K^+K^-$. 
These decay modes proceed with $s\bar{s}$-popping process and provide essential information to the weak decays of the charmed baryons.

Studies on the decays $\Xi_{c}^{0}\to\Xi^{0}\pi^0$, $\Xi^0\eta$ and $\Xi^0\eta^{\prime}$ were reported by Belle and Belle II~\cite{Belle:2024ikp}, giving $\mathcal{B}(\Xi_{c}^0\to\Xi^0\pi^0)=(6.9\pm0.3\pm0.5\pm1.3)\times10^{-3}$, $\mathcal{B}(\Xi_{c}^0\to\Xi^0\eta)=(1.6\pm0.2\pm0.2\pm0.3)\times10^{-3}$ and $\mathcal{B}(\Xi_{c}^{0}\to\Xi^0\eta^{\prime})=(1.2\pm0.3\pm0.1\pm0.2)\times10^{-3}$. 
Theoretical predictions based on the SU(3) breaking model~\cite{Zhong:2022exp} are consistent with the determined BFs for these $\Xi_{c}^{0}$ decays.

Belle also measured the BFs of the decays $\Xi_{c}^+\to\Sigma^+K^0_{S}$ and $\Xi^0\pi^+$ taking the decay $\Xi_{c}^{+}\to\Xi^-\pi^+\pi^+$ as the reference channel \cite{Belle-II:2025klu}, which gives $\mathcal{B}(\Xi_{c}^+\to\Sigma^+K^0_{S})=(0.194\pm0.021\pm0.009\pm0.087)\%$ and $\mathcal{B}(\Xi_{c}^+\to\Xi^0\pi^+)=(0.719\pm0.014\pm0.024\pm0.322)\%$, respectively.
It is found that the result of $\mathcal{B}(\Xi_{c}^+\to\Sigma^+K^0_{S})$ is overall lower than most of the theoretical predictions.

On the aspects of CS $\Xcz$ decays, the first observations of such suppressed decays are $\Xi_{c}^0\to\Xi^-K^+$, $\Lambda K^+K^-$ and $\Lambda\phi$ by Belle~\cite{Belle:2013ntc}, which gave the BFs at the order of $10^{-4}$. See detailed values in Table~\ref{tab:XicHad_BF}.
The observed decay modes proceed through external and internal $W$-emission diagrams along with admixture of the $W$-exchange diagram.
These measurements can be used to study the corresponding decay dynamics and to investigate quantitatively the interplay between strong and weak interactions in charmed baryon weak decays.

For the charged $\Xi_{c}^+$ baryon, the first CS decay observed is $\Xi_{c}^+\to p K^- \pi^+$ in the SELEX experiment~\cite{SELEX:1999xrm}, wich is taken as golden channel for tagging the $\Xi_{c}^+$ in experiments. Later, additional CS decays of $\Xi_{c}^+\to \Sigma^+ \pi^+ \pi^-$, $\Sigma^- \pi^+ \pi^+$ and $\Sigma^+ K^+ K^-$ were observed by SELEX and FOCUS~\cite{ParticleDataGroup:2024cfk}. Recenlty, Belle and Belle II observed new CS decays of 
$\Xi_{c}^+\to p K^0_{S}$, $\Lambda\pi^+$, $\Sigma^0\pi^+$ and $\Xi^0 K^+$. Their BFs are measured to be $\mathcal{B}(\Xi_{c}^+\to pK^0_{S})=(7.16\pm0.46\pm0.20\pm3.21)\times10^{-4}$, $\mathcal{B}(\Xi_{c}^+\to \Lambda\pi)=(4.52\pm0.41\pm0.26\pm2.03)\times10^{-4}$, $\mathcal{B}(\Xi_{c}^+\to \Sigma^0\pi^+)=(1.20\pm0.08\pm0.07\pm0.54)\times10^{-3}$ and $\mathcal{B}(\Xi_{c}^+\to \Xi^0K^+)=(0.49\pm0.07\pm0.02\pm0.22)\times10^{-3}$ \cite{Belle-II:2025bgw,Belle-II:2025klu}, where
the dominate uncertainties come from the uncertainty of the reference channel $\Xi_{c}^{0}\to\Xi^-\pi^+\pi^+$.
Belle II also searched for CP violation in the CS decays of $\Xcp\to\Sigma^+ K^+ K^-$ and $\Sigma^+ \pi^+ \pi^-$, which gives $A^{\Xcp}_{CP}(\Sigma^+ K^+ K^-)=(3.7\pm6.6\pm0.6)\%$ and $A^{\Xcp}_{CP}(\Sigma^+ \pi^+ \pi^-)=(9.5\pm6.8\pm0.5)\%$~\cite{Belle-II:2025xvc}, respectively. Hence, no evidence of CP violation is found in these decays.
In addition, LHCb observed the first doubly CS decay of the $\Xi_c$ baryon, $\Xi_c^+\to p\phi$ via $\phi\to K^+K^-$, with a statistical significance of more than $15\sigma$~\cite{LHCb:2019nxp}. The BF is largely suppressed as $\BR(\Xi_c^+\to p\phi)=(1.2\pm0.6)\times 10^{-4}$.

In the golden channel of $\Xi_{c}^+\to pK^-\pi^+$, LHCb performed the first amplitude analysis based on about 133,000 signal events~\cite{LHCb:2025hul} originated form the semileptonic beauty-hadron decays. The PWA fit methodology basically follows the similar analysis of $\Lcp\to p K^- \pi^+$~\cite{LHCb:2022sck}. It is found that the most important intermediate resonances are the $\Kstarb(892)^0$, $\Delta(1232)^{++}$ and $\Kstarbzz(1430)^0$ states. Among the $\Lz^*$ resonances, the largest contribution is from the $\Lz(2000)$ state. The fit fractions and the derived BFs for the intermediate resonances are determined, as listed in Table~\ref{tab:XicHad_BF}.

Predominantly the $\Xi_{c}^{0}$ baryon decays into charmless final states via the $c \to s u\overline{d}$ transition. It can, however, also disintegrate into a $\pi^-$ meson and a $\Lambda_{c}^+$ baryon via $s$ quark decay or via $cs\to d c$ weak scattering.
In 2020, LHCb firstly observed this CS decay and determined the BF of $\Xi_{c}^{0}\to\Lambda_{c}^+\pi^-$ to be $(0.55\pm0.02\pm0.18)\%$ \cite{LHCb:2020gge}.
In 2023, Belle confirmed the discovery of $\Xi_{c}^0\to\Lambda_c^+\pi^-$ with $\mathcal{B}(\Xi_{c}^0\to\Lambda_c^+\pi^-)=(0.54\pm0.05\pm0.05\pm0.12)$~\cite{Belle:2022kqi}, which is consistent with the LHCb measurement.
The averaged $\mathcal{B}(\Xi_{c}^0\to\Lambda_c^+\pi^-)$ is basically larger than different theoretical predictions, which are all below 0.4\%~\cite{Cheng:1992ff, Voloshin:2019ngb, Gronau:2016xiq,Faller:2015oma,Cheng:2015ckx}.
This discrepancy would improves our understanding of the charmed baryon decays and the underlying dynamics of the weak decays. For example, a modified treatement in the bag model from the past version~\cite{Cheng:2015ckx} is implemented in order to match the experimental result~\cite{Cheng:2022jbr}.


Recent measurements of the polarization parameters in the decays of $\Xi_c$ are summarized in Table~\ref{tab:XicHad_PO}.
The polarization parameters of the neutral $\Xcz$ baryon are firstly determined in several decay modes, including $\Xcz\to \Xi^-\pi^+$, $\Xi^0\pi^0$, $\Lambda \bar{K}^*(892)^0$ and $\Sigma^+ K^*(892)^-$.
In 2001, the first measurement of the $\Xi^0_c$ polarization parameter in the decay $\Xi^0_c\to \Xi^- \pi^+$ was firstly reported as $\alpha^{\Xcz}_{\Xi^-\pi^+}=-0.56\pm0.39^{+0.10}_{-0.09}$ by CLEO\cite{CLEO:2000lsg}, and BELLE updated the result as $\alpha^{\Xcz}_{\Xi^-\pi^+}=-0.63\pm0.03\pm0.01$  with precision improved by more than one order of magnitude~\cite{Belle:2021crz}.
In the same year, Belle reported the polarization parameters, $\alpha^{\Xcz}_{\Lambda\bar{K}^*(892)^0}=0.15\pm0.22\pm0.04$ and $\alpha^{\Xcz}_{\Sigma^+K^*(892)^-}=-0.52\pm0.30\pm0.02$~\cite{Belle:2021zsy}, where the $\bar{K}^*(892)^0$ and $K^*(892)^-$ are reconstructed in the $K^-\pi^+$ and $K^0_S\pi^-$ final states, respectively. 
Based on the Belle and Belle II data, the first determination of the $\Xcz$ polarization parameter in $\Xcz\to \Xi^0\pi^0$ was implemented in 2024, which gave $\alpha^{\Xcz}_{\Xi^0\pi^0}=-0.90\pm0.15\pm0.23$~\cite{Belle:2024ikp}. 
As theoretical predictions on $\alpha^{\Xcz}_{\Xi^0\pi^0}$ span a wide range from $-1.0$ to $0.94$~\cite{Korner:1992wi,Ivanov:1997ra,Xu:1992vc,Cheng:1993gf,Zenczykowski:1993jm,Zou:2019kzq,Sharma:1998rd,Geng:2018plk,Geng:2019xbo,Zhao:2018mov,Huang:2021aqu,Hsiao:2021nsc,Zhong:2022exp,Xing:2023dni,Geng:2023pkr,Zhong:2024qqs}, 
the Belle~(II) result provide valuable input for refining different effective models in charmed baryon decays.
Only in 2025, the first measurement of the polarization parameters in the $\Xi_c^+$ decays is implemented by LHCb via the PWA of $\Xi_c^+\to pK^-\pi^+$, which gives the overall polarization parameter  $\alpha^{\Xcp}_{pK^-\pi^+}=0.691\pm0.005\pm0.030$~\cite{LHCb:2025hul}.
In the study, many non-zero polarization parameters are identified, such as $\alpha^{\Xcp}_{p\bar{K}^*(892)^0}=0.613\pm0.065$, $\alpha^{\Xcp}_{p\bar{K}^*_0(1430)^0}=-0.76\pm0.10$, $\alpha^{\Xcp}_{\Lambda(1520)\pi^+}=-0.77\pm0.13$ and $\alpha^{\Xcp}_{\Delta(1232)^{++}K^-}=-0.774\pm0.071$, indicating parity violation in these resonant decay contributions.
Currently the  $\Xi_c$ polarization parameters are much less studied compared to those of the $\Lcp$, which  underscores the need for systeamtic experimental efforts in order to constrain the underlying strong and weak dynamics governing the decays of the chamred bayrons.

\begin{table*}[tbp]
  \caption{The determined polarization parameters in $\Xi^0_c$ decays.} \label{tab:XicHad_PO}
  \begin{center}
    \begin{tabular}{lcc|lcc}
      \hline\hline
      Mode                                 & polarization $\alpha$ & Experiment                   &      Mode                                 & polarization $\alpha$ & Experiment                         \\
      \hline
      $\Xi^0_c\to \Xi^-\pi^+$              & $-0.63\pm0.03$        & Belle(2021)\cite{Belle:2021crz}     &     $\Xi^0_c\to \Lambda\bar{K}^*(892)^0$ & $0.15\pm0.22$         & Belle(2021)\cite{Belle:2021zsy}     \\
      $\Xi^0_c\to \Xi^0\pi^0$              & $-0.90\pm0.27$        & Belle(II)(2024)\cite{Belle:2024ikp} & 
      $\Xi^0_c\to \Sigma^+ K^*(892)^-$     & $-0.52\pm0.30$        & Belle(2021)\cite{Belle:2021zsy}     \\
      \hline \hline
    \end{tabular}
    \begin{tabular}{lc|lc|lc}
\multicolumn{6}{c}{Polarization parameter $\alpha$ in $\Xi^+_c$ decays from PWA of $\Xi^+_c\to pK^-\pi^+$ at LHCb~\cite{LHCb:2025hul} } \\ \hline
Decay & polarization $\alpha$ & Decay & polarization $\alpha$ & Decay & polarization $\alpha$ \\ \hline
$p\bar{K}^{*}(892)^{0}$ & $0.613 \pm 0.065$ &
$p\bar{K}_{2}^{*}(1430)^{0}$ & $0.36 \pm 0.17$ &
$p\bar{K}_{0}^{*}(700)^{0}$ & $0.60 \pm 0.12$ \\
$p\bar{K}_{0}^{*}(1430)^{0}$ & $-0.76 \pm 0.10$ &
$\Lambda(1405)\pi^+$ & $-0.75 \pm 0.30$ &
$\Lambda(1520)\pi^+$ & $-0.77 \pm 0.13$ \\
$\Lambda(1600)\pi^+$ & $-0.06 \pm 0.41$ &
$\Lambda(1670)\pi^+$ & $-0.66 \pm 0.19$ &
$\Lambda(1690)\pi^+$ & $-0.58 \pm 0.16$ \\
$\Lambda(1710)\pi^+$ & $-0.86 \pm 0.36$ &
$\Lambda(1800)\pi^+$ & $-0.5 \pm 1.2$ &
$\Lambda(1810)\pi^+$ & $0.96 \pm 0.43$ \\
$\Lambda(1820)\pi^+$ & $0.64 \pm 0.33$ &
$\Lambda(1830)\pi^+$ & $0.30 \pm 1.02$ &
$\Lambda(1890)\pi^+$ & $-0.19 \pm 0.58$ \\
$\Lambda(2000)\pi^+$ & $0.53 \pm 0.15$ &
$\Delta(1232)^{++}K^-$ & $-0.774 \pm 0.071$ &
$\Delta(1600)^{++}K^-$ & $0.35 \pm 0.28$ \\
$\Delta(1620)^{++}K^-$ & $0.26 \pm 0.39$ &
$\Delta(1700)^{++}K^-$ & $0.15 \pm 0.30$ \\
\hline \hline
    \end{tabular}
  \end{center}
\end{table*}

\subsection{$\Omega^0_c$ decays}

Currently, the absolute BFs of the $\Omega_{c}^{0}$ decay have not been determined in experiment.
The relative BFs of the $\Omega_{c}^{0}$ decays were measured taking the decay $\Omega_{c}^{0}\to\Omega^{-}\pi^{+}$ as the reference channel, as listed in Table~\ref{tab:OcHad_BF}.
In 2018, Belle reported the measurements of the relative BFs of the decays $\Omega_{c}^{0}\to\Omega^-\pi^+\pi^0$, $\Omega^-\pi^+\pi^-\pi^+$, $\Xi^-K^-\pi^+\pi^+$ and $\Xi^{0}K^-\pi^+$, and the first measurements of the BFs of the decays $\Omega_{c}^{0}\to\Xi^-\bar{K}^0\pi^+$, $\Xi^0\bar{K}^0$ and $\Lambda\bar{K}^0\bar{K}^0$ based on the full Belle data~\cite{Belle:2017szm}.
The Belle measurements of these decay rates have provided substantial improvement in  precisions.
In addition, resonant states of the $\Xi^0(1530)$ and $\bar{K}(892)^{*0}$ were found to be prominent with fractions of about 33\% and 55\%, respectively, in the decay of $\Ocz\to \Xi^- K^-\pi^+\pi^+$, the $\rho(770)^+$ state takes fraction of larger than 71\% in the decay $\Ocz\to \Omega^-\pi^+\pi^0$, and the $\bar{K}^{*}(892)^0$ resonance contributes about 57\% in the decay $\Ocz\to \Xi^0K^-\pi^+$.
Furthermore, evidence for the $\Omega(2012)^-$ in the decays $\Omega_{c}^{0}\to \Xi^-  \bar{K}^0 \pi^+$ and $\Xi^0 K^- \pi^+$, where $\Omega(2012)^- \to \Xi^0 K^- $ and $\Xi^-\bar{K}^0$~\cite{Belle:2021gtf}.

\begin{table*}[tph]
  \caption{The measured relative BFs for the $\Omega^0_c$ decays with respect to the mode $\Omega^0_c\to\Omega^-\pi^+$.  Upper limits are set at 90\% confidence level.} \label{tab:OcHad_BF}
  \begin{center}
    \begin{tabular}{lcc|lcc}
      \hline\hline
      Mode                                      & reltive BF         & Experiment                      & Mode                                   & reltive BF         & Experiment                      \\
      \hline
      $\Omega^0_c\to\Xi^0\bar{K}^0$             & $1.64\pm0.29$ & Belle(2018)\cite{Belle:2017szm} & $\Omega^0_c\to\Sigma^+K^-K^-\pi^+$     & $<0.32$       & Belle(2018)\cite{Belle:2017szm} \\
      \hline
      \multirow{2}*{$\Omega^0_c\to\Xi^-\pi^+$}  & $0.25\pm0.06$ & Belle(2023)\cite{Belle:2022yaq} & 
      $\Omega^0_c\to\Lambda\bar{K}^0\bar{K}^0$  & $1.72\pm0.35$  & Belle(2018)\cite{Belle:2017szm} \\ \cline{4-6}
      ~  & $0.16\pm0.01$ & LHCb(2024)\cite{LHCb:2023fvd}   & $\Omega^0_c\to\Xi^0K^-\pi^+$           & $1.20\pm0.18$ & Belle(2018)\cite{Belle:2017szm} \\
      \hline
      $\Omega^0_c\to\Xi^-K^+$                   & $<0.07$       & Belle(2023)\cite{Belle:2022yaq} & $\Omega^0_c\to\Xi^-\bar{K}^0\pi^+$     & $2.12\pm0.28$ & Belle(2018)\cite{Belle:2017szm} \\
      \hline
      \multirow{2}*{$\Omega^0_c\to\Omega^-K^+$} & $<0.29$       & Belle(2023)\cite{Belle:2022yaq} & $\Omega^0_c\to\Xi^-K^-\pi^+\pi^+$      & $0.68\pm0.08$ & Belle(2018)\cite{Belle:2017szm} \\ \cline{4-6}
      ~                                         & $0.06\pm0.01$ & LHCb(2024)\cite{LHCb:2023fvd}   & $\Omega^0_c\to\Omega^-\pi^+\pi^0$      & $2.00\pm0.20$ & Belle(2018)\cite{Belle:2017szm} \\
      \hline
      & & & $\Omega^0_c\to\Omega^-\pi^+\pi^-\pi^+$ & $0.32\pm0.05$  & Belle(2018)\cite{Belle:2017szm} \\
      \hline \hline
    \end{tabular}
  \end{center}
\end{table*}

\begin{figure}[tbp]
  \begin{center}
    \includegraphics[width=0.325\linewidth]{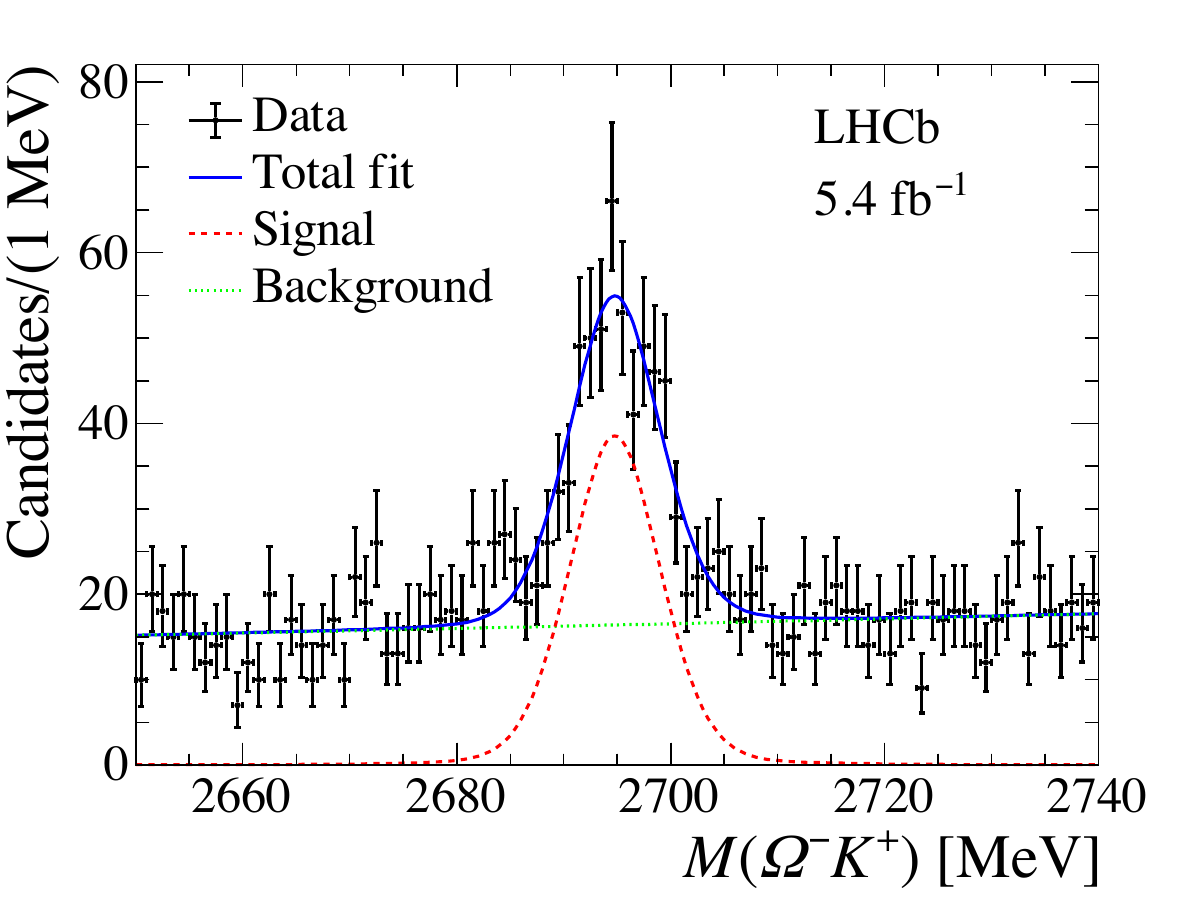}
    \includegraphics[width=0.325\linewidth]{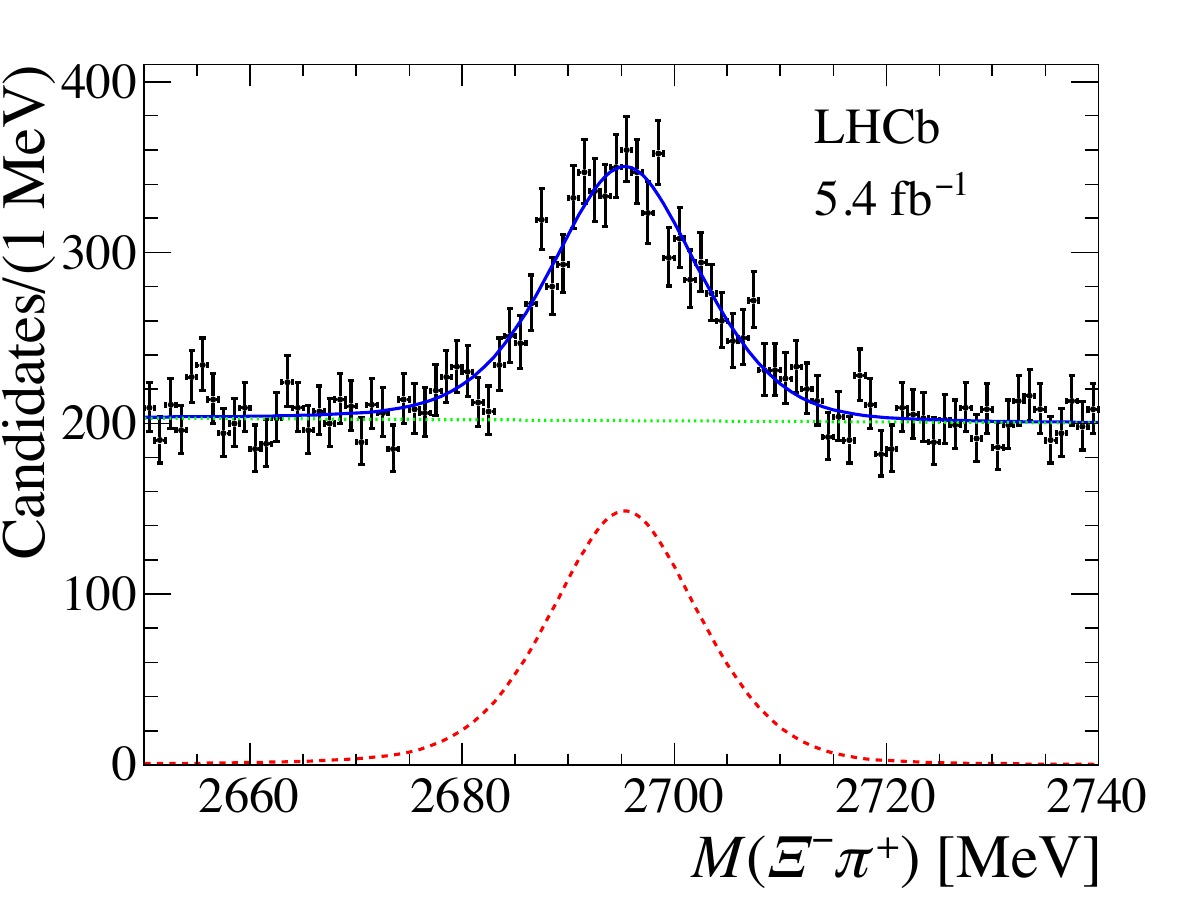}
    \includegraphics[width=0.325\linewidth]{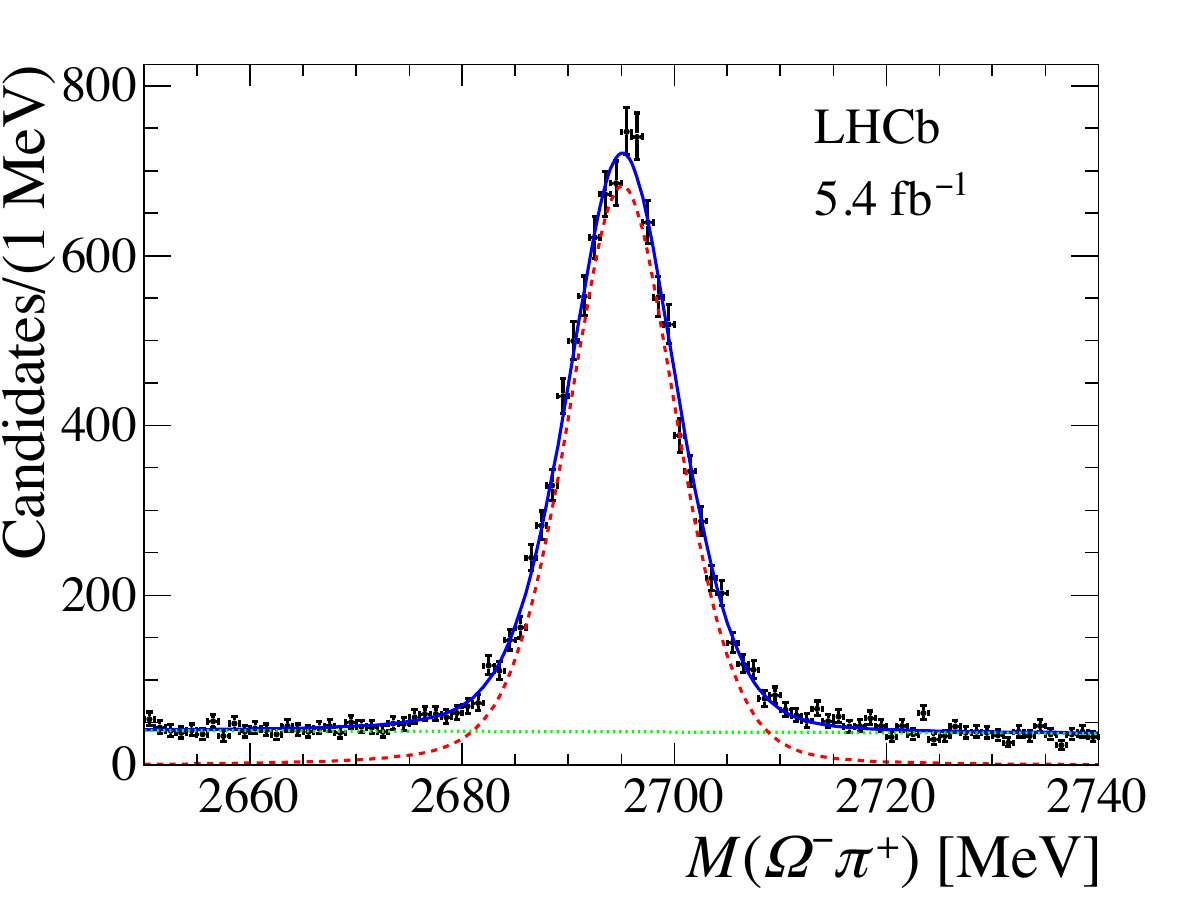}
    \vspace*{-0.5cm}
  \end{center}
  \caption{
    In the LHCb experiment, invariant mass distributions of the reconstructed $\Ocz$ candidates via (left) $\Omega_{c}^0\to\Omega^-\pi^+$, (middle) $\Omega_{c}^0\to\Xi^-\pi^+$, and (right) $\Omega_{c}^0\to\Omega^-\pi^+$ decays~\cite{LHCb:2023fvd}.}
  \label{fig:OmgC_results}
\end{figure}

For the suppressed decays, the first mode observed is $\Ocz\to  p K^- K^- \pi^+$ at LHCb, which is taken as tag mode for studies on the $\Ocz$~\cite{LHCb:2016coe,LHCb:2018nfa,LHCb:2021vll}. Although this channel has a small BF, it has advantage of a larger acceptance in the LHCb detector compared to decay modes with hyperons in the final states. However, no BF measurement is available yet for this decay.
Belle for the first time studied the SCS decays $\Omega_{c}^0\to\Xi^-\pi^+$ and $\Omega_{c}^0\to\Omega^-K^+$, as well as the DCS decay $\Omega_{c}^0\to\Xi^-K^+$~\cite{Belle:2022yaq}. An evidence of $\Omega_{c}^0\to\Xi^-\pi^+$ is seen with a BF of about one fourth of that for $\Omega_{c}^{0}\to\Omega^{-}\pi^{+}$, and no significant signals were found for the decays $\Omega_{c}^0\to\Omega^-K^+$ and $\Xi^-K^+$.
Based on proton-proton collision data at 13\,TeV taken between 2016 and 2018, corresponding to a luminosity of 5.4\,\ifb, LHCb firstly observed the SCS decays $\Omega_{c}^0\to\Omega^-K^+$ and $\Xi^-\pi^+$~\cite{LHCb:2023fvd}, where their relative BFs are measured to be at least one order of magnitude less than that for the reference mode $\Omega_{c}^{0}\to\Omega^{-}\pi^{+}$.
Figure~\ref{fig:OmgC_results} shows the invariant mass distributions of the signal modes of $\Omega_{c}^0\to\Omega^-K^+$ and $\Omega_{c}^0\to\Xi^-\pi^+$, and the reference mode $\Omega_{c}^0\to\Omega^-\pi^+$. 
At the same time, the most precise measurement of  for the $\Omega_{c}^0$ mass was carried out based on fit to the invariant mass distributions in Fig.~\ref{fig:OmgC_results}. 

The above BF measurements of the $\Ocz$ decay rates supply important information for the understanding of the weak decays of the charmed baryons.
Although there are no experimental measurements of the polarizaiton parameters of the $\Omega_{c}^0$ decyas, it is foreseen that BESIII, LHCb and Belle (II) have great opportunities to implement those studies in the near future~\cite{Wang:2016elx}.

  \section{Rare decays}
  Weak radiative decays, such as $\Lambda_c^+ \to \Sigma^+ \gamma$, $\Xcz \to \Xi^0 \gamma$, $\Lambda_c^+ \to p \gamma$, $\Xi_c^{+,0} \to \Sigma^{+,0} \gamma$ and $\Xi_c^{0} \to \Lambda \gamma$, can occur through bremsstrahlung processes in $W$-exchange process. 
The former two modes proceeds via the Cabibbo-favored transition $cd \to us\gamma$, whereas the latter processes involve the Cabibbo-suppressed transition $cd \to ud\gamma$ or $cs \to us\gamma$. 
In SM, the BFs for the Cabbibo-favored modes are estimated to be on the order of $10^{-4}\sim10^{-5}$~\cite{Kamal:1983zt,Uppal:1992cc,Cheng:1994kp,Shi:2022zzh}, while those for the Cabbibo-suppressed modes are at least one order of magnitude lower~\cite{Adolph:2022ujd}.
In experiment, $\Lambda_c^+ \to \Sigma^+ \gamma$ are searched for at BELLE~\cite{Belle:2022raw} and BESIII~\cite{BESIII:2022aok}, and the upper limit of the BF at 90\% confidence level is obtained to be $2.6\times 10^{-4}$. 
BELLE also reported the UL of $\BR(\Xcz \to \Xi^0 \gamma)$ as $1.8\times 10^{-4}$ at 90\% confidence level~\cite{Belle:2022raw}. 
The two ULs are at the order of some theoretical estimations.

Flavor Changing Neutral Current (FCNC) decays of $c \to u \ell^+ \ell^-$, are prohibited at the tree level and occur only through loop diagrams. In charm sector, this short distance contribution is expected to be extremely small With BF below $\mathcal{O}(10^-8)$, due to the severe suppression by the Glashow-Iliopoulos-Maiani mechanism in SM.
However, the BF can be enhanced by long-distance dynamics in SM to the order of $10^{-6}$. The FCNC decays of the charmed meson $D^0 \to \pi^+ \pi^- \mu^+ \mu^-$, $D^0 \to K^+ K^- \mu^+ \mu^-$, and $D^0 \to K^- \pi^+ \mu^+ \mu^-$ have been observed at LHCb with BFs at the $10^{-7}$ level~\cite{ParticleDataGroup:2024cfk}, which suggest non-trivial contributions from complicated long-distance effects. To date, in the charmed baryon sector, only the decay $\Lcp \to p \ell^+ \ell^-$ has been searched for and the best UL is obtained by LHCb with $\BR(\Lcp \to p \mumu)<2.9\times 10^{-8}$ at 90\% confidence level~\cite{LHCb:2024hju}, which suppressed the previous limits set by E653, Babar and LHCb~\cite{E653:1995rpz,BaBar:2011ouc,LHCb:2017yqf}. 
In addition, the ULs of the BFs for $\Xcz\to \Xi^0 \ee$ $\Xi_c\to \Xi \mumu$ are given at BELLE~\cite{Belle:2023ngs} as $9.9\times 10^{-5}$ and $6.5\times 10^{-5}$, respectively.

Although the SM-allowed FCNC decay rates are small, the new physics effects, such as minimal supersymmetric SM with R-parity violation and the two-Higgs-doublet model, can enhance the BFs by more than two orders of magnitude. 
Hence, experimental stuides of more FCNC decays of the charmed baryons serve as sensitive probes to test SM and search for new physics~\cite{Zhang:2024asb}.
For instance, BESIII performed search for exotic massless dark photon $\gamma^\prime$ in new physics model via the rare FCNC decay $\Lcp \to p \gamma^\prime$, which gives $\BR(\Lcp\to p \gamma^\prime)<8\times 10^{-5}$ at 90\% conference level~\cite{BESIII:2022vrr}.

  \section{Future aspects}
  Similar to the experimental situation of the lightest charmed baryon, $\Lcp$, prior to 2014, current data on the $\Xi_c$ and $\Ocz$ remain very limited. Only a few decay modes have been observed, and the measured BFs carry large uncertainties. To date, there are no absolute BF measurements for the $\Ocz$ baryon. 
The absolute BFs for the $\Xcz$ are known with poor precision, uncertainties exceeding 20\%, and those for the $\Xcp$ are even less precise, with uncertainties greater than 40\%. So far, only one doubly CS decay of the charmed baryon, $\Lcp \to p K^+ \pi^-$, has been observed~\cite{Belle:2015wxn,LHCb:2017xtf}. More precise studies of additional CS decays of charmed baryons are highly desirable in the future. 
In addition, so far only a few measurements of decay asymmetries in $\Xcz$ decays were implemented, and no decay asymmetries of $\Xcp$ and $\Ocz$ are reported. 
Further efforts of such studies are crucial not only for improving our knowledge on strong dynamics in charm region, but also for the searches for CP violation in charmed baryons, which has not yet been observed.

The experimental observables of the decay BF and polarization parameter are only resembles of the invovled weak transition diagrams and complicated non-perturbagtive strong dynamics in the process.
The long-distance contributions~\cite{Jia:2024pyb,Cheng:2025oyr,Yang:2025orn}, such as the final state interactions, play important roles in the hadronic decays of charmed baryons. 
That means a systematic measurement is needed to calibrate the contributions of different ingredients.
This has been demonstrated in the studies of charmed mesons, in which systematic precision measurements of different types of the hadronic BFs for the charmed mesons at CLEO and BESIII, including the CF, CS, and DCS decays, greatly improved the prediction power in the CP violation of charmed mesons in theory~\cite{Saur:2020rgd}. Following the similar strategy, the future comprehensive measurements of the hadronic BFs, as well as the polarization asymmetries, of the charmed baryons will provide important constraints to the theoretical models, and then significantly improve the model uncertainties on the CP violation of charmed baryons~\cite{Cheng:2025oyr}.

With advancement of experimental studies on the $\Lcp$ decays, we see that SU(3) flavor symmetry breaking became more and more evident in the $\Lcp$ decays~\cite{Geng:2025wfi,Yang:2025orn}. The sizes of the SU(3) sysmmetry breaking provide crucial information to understand the involved dynamics related to the SU(3) flavor symmetry breaking in the charmed baryons.
In the coming years, thorough experimental tests on the SU(3) breaking effect in the $\Xi_c$ and $\Ocz$ decays will be an important task, in order to map out a whole picture of the SU(3) flavor symmetry breaking in the charmed baryons. 

These efforts will provide important constraints on the theoretical predictions on the CP violation in charmed baryons. In SM, the amplitude of the CP violation in charmed baryons are predicted at the order of 0.1\%~\cite{Bigi:2012ev}. 
However, the experimental searches for CP violation have not yet yielded conclusive results. 
For three-body decays, the LHCb measured the difference in the CP asymmetries of $\Lambda_c^+ \to p K^+ K^-$ and $\Lambda_c^+ \to p \pi^+ \pi^-$, $\Delta A_{CP}$, as $(0.30 \pm 0.91 \pm 0.61)\%$~\cite{LHCb:2017hwf}, as well as the CP asymmetry in $\Xcp \to p K^- \pi^+$~\cite{LHCb:2020zkk}.
These results are  consistent with SM preditions with limitation of the statistical uncertainties.
An increase in statistics by at least a factor of 100 is required to probe the size of the CP asymmetry in SM.
The four-body final states in hadronic decays, such as $\Lambda_c^+\to p\pi^+\pi^-\pi^0$, $\Lambda_c^+\to\Lambda K^+ \pi^+\pi^-$, and $\Lambda_c^+\to pK_S\pi^+\pi^-$ and $\Xcz \to p K^- \pi^+\pi^-$, offer opportunities to explore CP violations through T-odd observables.
Driven by the increasing samples of the charmed baryons, especially for the $\Lcp$, collected in the BESIII, LHCb, and Belle II experiments, more searches for CP violations in charmed baryons are expected. 

LQCD is a powerful tool for studying non-perturbative QCD effects in hadrons. In the charm sector, LQCD has made significant progress in calculating the properties of charmed mesons, such as decay constants and form factors, with high precision~\cite{FlavourLatticeAveragingGroupFLAG:2024oxs}. However, there meets still challenges in extending these calculations to charmed baryons due to their more complex structure.  The SL decays of charmed baryons provides a clean environment to disentangle strong and weak interactions, enabling the extraction of form factors for comparison with LQCD calculations. 
Recent LQCD calculations on $\br{\Xcz \to \Xi^- e^+ \nu_e}$~\cite{Farrell:2025gis,Zhang:2021oja} show disagreements between each other, and both predictions are significantly higher than the Belle  measurement~\cite{Belle:2021crz}.
In experiment, except for the $\Lcp \to \Lambda \ell^+ \nu_\ell$ decay~\cite{BESIII:2019odb}, form factor measurements in SL decays have not yet been experimentally studied. 
In addition, CS SL decays of the $\Xi_c$ and $\Ocz$ baryons remain unexplored.
To address these challenges, threshold production of charmed baryons at experiments such as BESIII and the proposed Super $\tau$-Charm Factory (STCF)~\cite{Achasov:2023gey} offers a unique and essential opportunity.
This would allow systemactic studies of the properties of all the singly charmed baryons, which are still lacking, in order to supply thorough comparisons and calibrations with the LQCD caculations.

\begin{figure}[tp]
  \centering
  \includegraphics[width=0.45\textwidth]{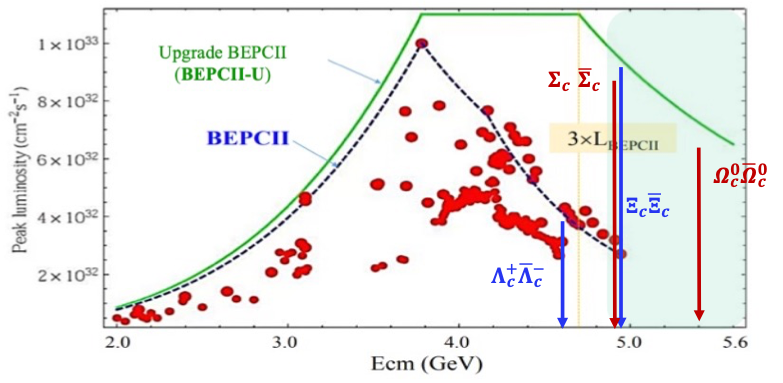} \hspace{0.1cm}
  \includegraphics[width=0.45\textwidth]{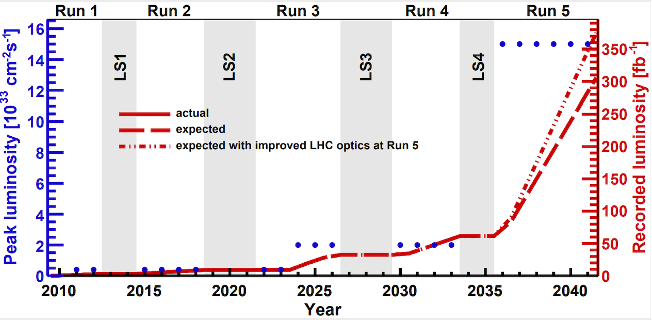} \\
  \includegraphics[width=0.4\textwidth]{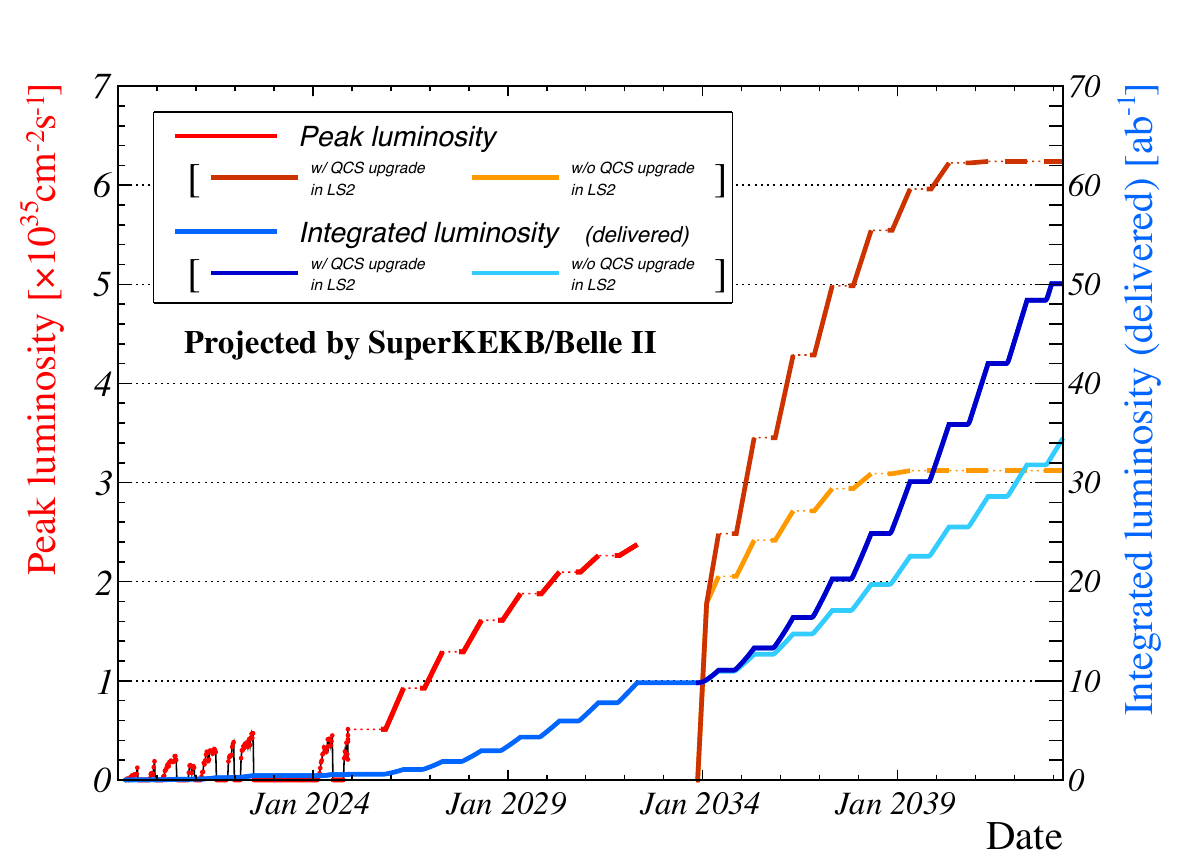}
  \caption{The current and future propects of BESIII (upper left), LHCb (upper right) and BELLE (II) (bottom) physics runs.
  \label{fig:future_runs}}
  \end{figure}

With the quantum correlation of the spin-half charmed baryons in threshold pair production, BESIII will offer a significant enhancement in the sensitivity of decay asymmetries and searches for CP violation by combining ``single tag" $\Lambda_c^+$ data~\cite{BESIII:2019odb} with ``double tag" $\Lambda_c^+\bar{\Lambda}_c^-$ data, where the pairs of $\Lambda_c^+\bar{\Lambda}_c^-$ are quantum-correlated with respect to their spins aligned to the initial transverse polarization of the virtual photon. In addition, using longitudinal polarized beam, the STCF can achieve enhanced sensitivities to decay asymmetries and CP violations, given the known direction of the spin orientation of the produced $\Lambda_c^+$~\cite{Wang:2025yjs}.


In the energy region just above the charmed baryon threshold, BESIII has unique advantage in extensive studies of the production and decay properties of (excited) charmed baryons. 
The BEPCII team has implemented a new accelerator upgrade plan, BEPCII-U, aiming to increase the maximum energy to 5.6 GeV, covering the production thresholds of all ground-state charmed baryons, as depicted in Fig.~\ref{fig:future_runs}. 
Simultaneously, the peak instantaneous luminosity at 4.7~GeV will be tripled, maintaining the highest instantaneous luminosity in the energy region from 4.6~GeV to 4.7~GeV energy range at $1.1 \times 10^{33}$ cm$^{-2}s^{-1}$
This upgrade significantly enhances the data-taking efficiency of BESIII in the charmed baryon energy region, ensuring to accumulate at least 18$\,\ifb$
of data in the energy range of 4.6$-$4.95~GeV proposed in the BESIII physics white paper~\cite{BESIII:2020nme}, which corresponds to approximately 2.5 million $\Lcp$  pairs.
Absolute measurements of the SL and non-leptonic decays of the $\Lambda_c^+$, $\Xi_c^{+,0}$, and $\Omega_c^0$ baryons will be significantly improved, including the decay asymmetries (such as $\alpha$) in various charmed baryon decays. The absolute BFs of $\Xi_c$ and $\Omega_c^0$ decays will also be measured. Moreover, searches for rare and forbidden decays of charmed hadrons could yield sensitivity improvements of up to two orders of magnitude, further advancing our quest for new physics.


LHCb has published a series of results on the properties of charmed baryons based on RUN1 and RUN2 data. 
As shown in Fig.~\ref{fig:future_runs}, the third operational period (RUN3), which began in 2023, has already gathered approximately 8.4$\,\ifb$ of data at 13~TeV and is expected to collect an additional ~17$\,\ifb$ during 2025-2026. 
The upgraded LHCb detector with enhanced components and improved trigger efficiencies~\cite{LHCb:2023hlw} are expected to increase the statistical sample size by a factor of ten compared to previous RUN1 and RUN2 data. This advancement will significantly refine the precision of charmed baryon studies.

So far the Belle~II experiment has accumulated 575$\,\ifb$ of data at the $\Upsilon(4S)$ resonance and in total about 5~ab$^{-1}$ of data will be accumulated by 2030. The new BELLE II detector is equipped with better resolution and particle identification, which enhances capability to cope with higher backgrounds. Therefore, the BELLE~II experiment is expected to provide a wealth of data on the properties of charmed baryons.

Therefore, it is foreseen that great progresses in experimental studies of charmed baryons will continue to be available in the coming years. Especially, with the realization of BEPCII-U, an comprehensive exploration on the $\Xi_c$ and $\Ocz$ properties 
will become feasible, similar to the previous efforts that enriched the experimental data on the $\Lcp$.


  \vspace{-1mm}
  \centerline{\rule{80mm}{0.1pt}}
  \vspace{2mm}

  \begin{multicols}{2}
    \bibliographystyle{apsrev4-2}
    \bibliography{refs}

  \end{multicols}

  \clearpage
\end{CJK*}
\end{document}